\documentclass[10pt,twocolumn,aps,pra,superscriptaddress,amsmath,showpacs,tightenlines,pdflatex,longbibliography,nofootinbib]{revtex4-1}

\usepackage[T1]{fontenc}
\usepackage{amsthm}
\usepackage{amsmath,bm}
\usepackage{latexsym}
\usepackage{amsfonts}
\usepackage{amssymb}
\usepackage{color}
\usepackage{bbm,dsfont}
\usepackage{graphicx}
\usepackage{subfigure}
\usepackage[colorlinks=true,linkcolor=violet, citecolor=cyan,urlcolor=magenta]{hyperref}
\usepackage{tikz}

\usepackage{times} 

\usepackage[normalem]{ulem}

\usepackage{ucs}

\AtBeginDocument{%
    \newwrite\bibnotes
    \def\bibnotesext{Notes.bib}
    \immediate\openout\bibnotes=\jobname\bibnotesext
    \immediate\write\bibnotes{@CONTROL{REVTEX41Control}}
    \immediate\write\bibnotes{@CONTROL{%
    apsrev41Control,author="08",editor="1",pages="1",title="0",year="1"}}
     \if@filesw
     \immediate\write\@auxout{\string\citation{apsrev41Control}}%
    \fi
}%

\usepackage{cleveref}
\crefname{equation}{Eq.}{Eqs.}
\crefname{section}{Sec.}{Sections}
\crefname{figure}{Fig.}{Figs.}

\newcommand{\id}{\mathbbm{1}} 

\newcommand{\Q}{\mathsf Q}

\renewcommand{\P}{\mathsf P}
\renewcommand{\S}{\mathsf S}
\newcommand{\W}{\mathsf W}

\newcommand{\M}{\mathsf M}
\newcommand{\II}{\mathcal I}

\newcommand{\mean}[1]{\langle #1 \rangle}

\newcommand{\qon}{{\bf q}_{\rm 1\rightarrow n}}

\newcommand{\son}{{\bf s}_{\rm 1\rightarrow n}}

\newcommand{\bq}{\mathbf{q}}
\newcommand{\bs}{\mathbf{s}}
\newcommand{\bqp}{\mathbf{q'}}
\newcommand{\bsp}{\mathbf{s'}}

\DeclareMathOperator{\tr}{tr}

\newcommand{\ket}[1]{| #1 \rangle}

\newcommand{\ketbra}[1]{| #1 \rangle\langle #1 |}
\newcommand{\Tr}{\mathrm{Tr}}

\definecolor{mycolor}{HTML}{fcd492}

\begin{document}
\title{Leggett-Garg macrorealism and temporal correlations}

\author{Giuseppe Vitagliano}
\email{giuseppe.vitagliano@tuwien.ac.at}
\affiliation{Vienna Center for Quantum Science and Technology, Atominstitut, TU Wien,  1020 Vienna, Austria}

\author{Costantino Budroni}
\email{costantino.budroni@unipi.it}
\affiliation{Department of Physics ``E. Fermi'' University of Pisa, Largo B. Pontecorvo 3, 56127 Pisa, Italy}
\affiliation{Faculty of Physics, University of Vienna, Boltzmanngasse 5, 1090 Vienna, Austria}
\affiliation{Institute for Quantum Optics and Quantum Information (IQOQI), Austrian Academy of Sciences, Boltzmanngasse 3, 1090 Vienna, Austria}

\begin{abstract}  

Leggett and Garg formulated macrorealist models encoding our intuition on classical systems, i.e., physical quantities have a definite value that can be measured with minimal disturbance, and with the goal of testing macroscopic quantum coherence effects. The associated inequalities, involving the statistics of sequential measurements on the system, are violated by quantum-mechanical predictions and experimental observations. Such tests, however, are subject to loopholes: a classical explanation can be recovered assuming specific models of measurement disturbance. We review recent theoretical and experimental progress in characterizing macrorealist and quantum temporal correlations, and in closing loopholes associated with Leggett-Garg tests. Finally, we review recent definitions of nonclassical temporal correlations, which go beyond macrorealist models by relaxing the assumption on the measurement disturbance, and their applications in sequential quantum information processing.

\end{abstract}

\maketitle

\section{Introduction}

The question of whether quantum mechanics is compatible with the assumption that physical quantities have a definite value at each instant of time can be traced back to Heisenberg's argument about uncertainties for position and momentum \cite{Heisenberg:1927ZP,sep-qt-uncertainty}. In the same years, the question of whether quantum effects can be witnessed at the macroscopic level was addressed by Schr\"odinger~\cite{Schrodinger1935} in his famous cat thought experiment.

Leggett and Garg (LG) combined these intuitions into the notion of {\it macroscopic realism}, or simply {\it macrorealism} \cite{LeggettPRL1985,EmaryRPP2014}. In a macrorealist model, a macroscopic physical quantity is considered, e.g., the position of a massive object that is displaced over macroscopic distances, and it is assumed that this quantity has a definite value at each time and that it is possible to measure it with an arbitrary small disturbance on its subsequent dynamics. These assumptions give rise to a hidden variable model, similar to those introduced by Bell~\cite{Bell1964,nonloc_rev} and Kochen and Specker~\cite{kochen67,Context_review}, and open the possibility of subjecting macrorealist (MR) models to experimental tests. Such tests are based on Leggett-Garg inequalities (LGIs), namely, bounds on the observed statistics coming from sequential measurements on a physical system that are respected by MR models, but violated by quantum-mechanical predictions.

 Similarly to Bell tests~\cite{LarssonJPA2014}, Leggett-Garg tests are subject to loopholes~\cite{WildeMizel2012}, either due to practical reasons related to the realization of the experiment or to more fundamental ones.  A considerable effort has been devoted into closing such loopholes in experimental tests of LGIs in recent years. This resulted in a variety of approaches, both on the experimental side and on the theoretical analysis of the results.
 
The notion of macrorealism, involving sequences of measurements in time, has become the standard notion of nonclassical temporal correlations. However, the assumption that measurements cause no disturbance on the subsequent dynamics of the system is rather strong and it is explicitly violated in many experimental setups, even simply because of limitations in the noise reduction in the measurement apparatus. 

Moreover, there is an interest in developing notions of nonclassical temporal correlations that can be applied to the investigation of quantum advantages in information processing tasks. For instance, a classical device with memory, which is updated at each time steps, clearly violates the LG assumption of nondisturbing measurements. 
This stimulated several approaches to redefine the notion of nonclassical temporal correlations from an operational perspective  \cite{Zukowski2014, BudroniNJP2019, BrierleyPRL2015, Ringbauer2017}. In the light of this, the assumption of a nondisturbing measurement can be relaxed to that of a bounded memory, i.e., a finite number of internal states, for the physical system. Leggett-Garg nondisturbing measurements are recovered in the case of a single internal state~\cite{Zukowski2014, BudroniNJP2019}. Besides this, a variety of different assumptions have been employed also in the attempt of closing the loopholes in LG tests.

In this perspective paper, we aim to cover the recent developments on LG tests, including the theoretical and experimental efforts to close all the loopholes, as well as the more recent extensions of LG ideas on nonclassical temporal correlations to the investigation of quantum information processing tasks. In particular, we consider experiments that are not covered by the previous review of Emary {\it et al.} on LGIs~\cite{EmaryRPP2014}.

The paper is organized as follows. In \cref{sec:1}, we recall the basic definition of macrorealism and introduce the theoretical tools for characterizing the set of macrorealist correlations and the quantum correlations arising in the temporal scenario. In \cref{Sec:exp_tests}, we address the problem of loopholes in LG tests, the theoretical proposals to address them, and the recent experiments that have implemented them. In \cref{sec:finitestates}, we discuss operational notions of nonclassical temporal correlations that go beyond the LG proposal by relaxing the assumption of noninvasive measurements, as well as their applications in quantum information processing. Finally in \cref{sec:outlook}, we conclude discussing future directions in the research on LG tests and temporal correlations.

\section{Leggett-Garg macrorealism}\label{sec:1}

\subsection{Original formulation}\label{sec:1A}
Let us start with the basic definition of {\it macrorealism} introduced by Leggett and Garg~\cite{LeggettPRL1985}. A {\it macrorealist theory} is defined by two main assumptions:
\begin{itemize}
\item[(MRPS)] {\it Macroscopic realism (per se)}: The value of a macroscopic quantity $Q(t)$ is well defined at each time $t$.
\item[(NIM)] {\it Noninvasive measurability}: it is possible, in principle, to measure the quantity $Q(t)$ with an arbitrarily small perturbation of its subsequent dynamics.
\end{itemize}
The adjective {\it subsequent} in the NIM assumption, implicitly contains an assumption on the causality properties of the scenario sometimes denoted as {\it induction}, namely:
\begin{itemize}
\item[(IND)] {\it Induction}: The outcome of a measurement is not influenced by what will be measured on the system at a later time.
\end{itemize}
This assumption is usually taken for granted in most theoretical and experimental investigation of Leggett-Garg MR models, unless one is interested in exotic causal structures or even models with retrocausal influences \cite{Wharton-Argaman2020,LeiferPRSA2017}


\begin{figure}
\includegraphics[width=\linewidth]{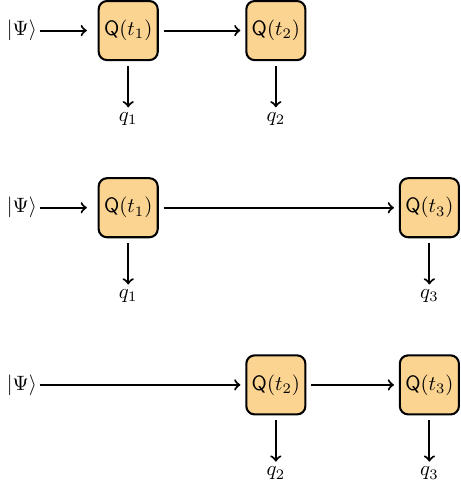}
\caption{Three-time measurement for the basic Leggett-Garg scenario. In each experimental run, the system is measured at two different time steps, $t_i$ and $t_j$, with $(i,j)=(1,2),(1,3)$ or $(2,3)$. The observed statistics is used to estimate the correlators $\mean{Q_iQ_j}$.}\label{fig:1}
\end{figure}


To derive the original LGI from these basic assumptions, we consider the measurement scenario in \cref{fig:1}. A measurement of the physical quantity $Q$ is performed sequentially for each possible pair of three time instants $t_1,t_2,$ and $t_3$, namely, at $(t_1,t_2)$,  $(t_1,t_3)$ or $(t_2,t_3)$. We use the convention of denoting by $Q$ the physical quantity under consideration, and by $\Q$ the associated quantum operator. The corresponding outcomes $q_i=\pm 1$, $i=,1,2,3$ are observed, giving rise to the statistical distributions
\begin{equation}
p_{12}(q_1,q_2), \ \ p_{13}(q_1,q_3),\ \ p_{23}(q_2,q_3).
\end{equation}

Under the assumption MRPS, at each time instant $t_i$ there exists a definite value $q_i$ for the physical quantity $Q(t_i)$. This implies that there exists a joint distribution $p_{123}(q_1,q_2,q_3)$ that describes the possible values of $Q$ during each experimental run. Moreover, the assumption NIM implies that the act of measurement does not change such a quantity. In other words, measuring a quantity and discarding the result is equivalent to not measuring it at all. In mathematical terms, this implies that the observed distributions $p_{ij}(q_i,q_j)$, obtained by measuring only at two time steps, are precisely the marginals of the global distribution $p_{123}(q_1,q_2,q_3)$, obtained by measuring at all times, namely
\begin{align}
\label{eq:marginals_a} p_{12}(q_1,q_2)=\sum_{q_3} p_{123}(q_1, q_2, q_3),\\
\label{eq:marginals_b} p_{13}(q_1,q_3)=\sum_{q_2} p_{123}(q_1,q_2,q_3),\\  
\label{eq:marginals_c} p_{23}(q_2,q_3)=\sum_{q_1} p_{123}(q_1,q_2,q_3).
\end{align}
Notice that the first condition, i.e., on the marginal $p_{12}$, automatically follows from the assumption IND, even if the measurement is actually invasive.

If a global distribution $p_{123}$ exists and the observations come from this distribution, one can verify that the observed correlators $\mean{Q_i Q_j} := \sum_{q_i q_j} q_i q_j p_{ij}(q_i, q_j)$ satisfy the condition
\begin{equation}\label{eq:LGI}
\mean{Q_1 Q_2} + \mean{Q_2 Q_3} -\mean{Q_1 Q_3} \leq 1.
\end{equation} 
A straightforward proof of the validity of Eq.~\eqref{eq:LGI} is obtained by noticing that 
\begin{equation}
\begin{split}
\mean{Q_1 Q_2} + \mean{Q_2 Q_3} -\mean{Q_1 Q_3} \\
= \sum_{q_1,q_2,q_3=\pm 1} (q_1 q_2 + q_2 q_3 -q_1 q_3) p(q_1,q_2,q_3)\\
\leq \max_{q_1,q_2,q_3} (q_1 q_2 + q_2 q_3 -q_1 q_3) = 1,
\end{split}
\end{equation}
where we used the fact that $p(q_1,q_2,q_3)\geq 0$ and $\sum_{q_1,q_2,q_3} p(q_1,q_2,q_3)=1$. It is clear, then, that the derivation of the LGI completely relies on the existence of a global distribution for $(q_1,q_2,q_3)$. This is the same assumption we find in the definition of Bell local theories~\cite{Bell1964,nonloc_rev} and Kochen-Specker noncontextual theories~\cite{kochen67, Klyachko:2008PRL, Kleinmann:2012PRL, Context_review}. As Avis {\it et al.} \cite{AvisPRA2010} noticed, this allows us to use powerful methods developed in those frameworks, such as the correlation polytope approach~\cite{Pitowsky:1989}, to analyze LGIs.

To conclude this section, we show the simplest violation of the inequality in \cref{eq:LGI} in a two-level system. Consider a qubit prepared in the state $\ket{0}$, and rotating in the $(y,z)$ plane, i.e., evolving according to the unitary $e^{-i\sigma_x t}$, on which we measure $\sigma_z$, where $\sigma_x,\sigma_y,$ and $\sigma_z$ are the Pauli matrices. In the Heisenberg picture, the rotated observables will be $\Q(t_i)= \mathbf{n_i}\cdot \bm{\sigma}$, where $\bm{\sigma}=(\sigma_x,\sigma_y,\sigma_z)$,  and the vectors $\mathbf{n_i}$ are rotated by $\pi/3$ at each step, i.e., $\mathbf{n_1}=(0,0,1)$, $\mathbf{n_2}=(0,\sin(\pi/3),\cos(\pi/3))$, and  $\mathbf{n_3}=(0,\sin(2\pi/3),\cos(2\pi/3))$. 
It can be easily shown that in this case of qubit observables, and more in general for dichotomic (i.e., two-outcome) observables, the correlators arising from the standard projective measurement, are calculated simply from the anti-commutator between the observables~\cite{FritzNJP2010}: $\mean{Q_i Q_j} = 
	\tfrac{1}{2} \mean{\{\Q(t_i) , \Q(t_j)\}} = \tfrac 1 2 \mathbf{n_i}\cdot \mathbf{n_j}$. Using this, it 
is straightforward to verify that 
\begin{equation}
\mean{Q_1 Q_2} + \mean{Q_2 Q_3} -\mean{Q_1 Q_3} = \frac{3}{2}> 1,
\end{equation}
thus giving  a violation of the macrorealist bound. 
Note that quantum mechanics allows for more general implementations of sequences of measurements, as we will explain in the following sections, which can also lead to stronger violations of the LGIs.

\subsection{Arrow-of-time and macrorealist polytopes}
In this section, we introduce a powerful method to study both (non)macrorealist correlations and more general temporal correlations, namely, a geometric description of such correlations in terms of convex polytopes~\cite{Grunbaum:2003}. This method has been developed by several authors for Bell inequalities \cite{Froissart:1981,GM:1984,Pitowsky:1986JMP,Pitowsky:1989} and explored by Avis {\it et al.} \cite{AvisPRA2010} in relation to LGIs. An extensive treatment of the macrorealist polytope and other related polytopes was given by Clemente and Kofler \cite{ClementePRL2016}.

It is convenient to first introduce some notation. Let us denote a sequence of measurement settings as $\mathbf{s}:=(s_1,\ldots,s_n)$. For a LG test, $s_i$ could be either $0$ (no measurement) or $1$ (measurement). Of course, nothing forbids us from considering more general measurement settings, but for the moment we restrict to this case to keep the notation simpler. Similarly, we denote by $\mathbf{q}=(q_1,\ldots, q_n)$ the sequence of measurement outcomes, e.g., $q_i=\pm 1$ for LG tests. For the case of $s_i=0$, i.e., no measurement, we follow the convention of \cite{ClementePRL2016} of assigning the outcome $0$ with probability $1$.

In the simplest LGI, we had $q_i=\pm 1$, but, again, more general situations can be considered. The observed correlations are, then, of the form
\begin{equation}
p(\mathbf{q}|\mathbf{s})=p(q_1,\ldots,q_n|s_1,\ldots,s_n),
\end{equation}
representing the probability of the sequence of outcomes $q_1,\ldots,q_n$ given the measurement settings $s_1,\ldots,s_n$. For instance, we have $p_{13}(q_1,q_3) = p(q_1,0,q_2|1,0,1)$ with the convention for the setting $s=0$ described above.

We now want to write the conditions on the correlations imposed by IND. To do so, it is convenient to introduce the following notation. We denote by $\qon$ the sequence of $q_i$ from $1$ to $n$, i.e., $\qon=(q_1,\ldots,q_n)$, and similarly for the settings $\son$.
With this, we can write the condition imposed by IND, also called {\it Arrow of Time} (AoT) constraints \cite{ClementePRL2016}. Notice that similar constraints are also defined in terms of {\it no backward in time signaling} (NBTS)~\cite{Guryanova2019}. For the case of Leggett-Garg tests, i.e., for settings $s_i=0,1$, they have the form
\begin{equation}\label{eq:IND}
\begin{split}
p(\qon| \son) = \sum_{q_{n+1}} p(\qon, q_{n+1}|\son, 1) \\
\text{ for all }  n.
\end{split}
\end{equation}
The above constraints, defined recursively for all $n$ and together with the positivity, i.e., $p(\qon| \son)\geq 0$, and normalization, i.e., $p(0|0)=1$, of probability, define a convex polytope, known as the {\it AoT polytope}~\cite{ClementePRL2016}. In fact, all constraints are given by linear equalities or inequalities.

This definition of the AoT polytope can be extended beyond the one needed in LG tests, to include more measurement settings for each time step. The AoT polytope can be thought of as an analog of the {\it nonsignaling} (NS) polytope~\cite{PopescuFPH1994} in Bell nonlocality~\cite{nonloc_rev}. As such, it represents correlations obtainable in classical, quantum, and more general probability theories, provided that the causality constraint of IND is satisfied. In fact, it can be shown that all points of the polytope are reachable by quantum strategies~\cite{FritzNJP2010} (see also the discussion in \cite{ClementePRL2016}) and even that all the extremal points of the polytope can be reached by classical strategies, i.e., satisfying the assumption of MR, but involving {\it invasive} measurements~\cite{HoffmannThesis2016,Hoffmann2018,AbbottPRA2016}. We provide a sketch of this argument and an extended discussion of AoT polytopes for more general measurement scenarios in \cref{sssec:TC_gen}

While AoT constraints are satisfied by any theory obeying causality, NIM imposes a stronger constraint on the set of possible correlations. The example we provided in Eqs.~\eqref{eq:marginals_b},\eqref{eq:marginals_c} can be generalized to arbitrary sequences as
\begin{equation}\label{eq:nsit_gen}
\begin{split}
&p({\bf q}_{\rm 1\rightarrow i-1},0,{\bf q}_{\rm i+1\rightarrow n}|{\bf s}_{\rm 1\rightarrow i-1},0,{\bf s}_{\rm i+1\rightarrow n})
\\ 
&=\sum_{q_{i}}p({\bf q}_{\rm 1\rightarrow n}|{\bf s}_{\rm 1\rightarrow n}),
\end{split}
\end{equation}
for all $i$, ${\bf q}_{\rm 1\rightarrow n}$ and ${\bf s}_{\rm 1\rightarrow n}$ with $(s_{i+1},\ldots,s_n) \neq (0,\ldots,0)$. Equation \eqref{eq:nsit_gen} encodes the fact that we cannot detect whether a measurement has been performed, and its outcome discarded, at some point in the measurement sequence. For this reason, it is called {\it no signaling in time} (NSIT) condition~\cite{KoflerPRA2013}; see also the formulation in Ref.~\cite{LiSR2012}.

\subsubsection{Differences between the temporal and spatial scenarios}
Notice that AoT and NSIT constraints recover the standard NS constraints. Of course, these conditions have a different interpretation in the temporal scenario and most of the intuition developed in the spatial scenario does not hold. In fact, while in the spatial scenario one is limited in the joint measurements that can be performed, there is no such restriction in the temporal scenario. To clarify this aspect, it is helpful to consider a basic example, i.e., the Clauser-Horne-Shimony-Holt~\cite{Clauser:1969PRL} (CHSH) Bell inequality and its temporal analog, i.e., the four-term LGI.

In the CHSH scenario, Alice can choose between two possible settings, i.e., $x=0,1$ corresponding to local measurements $A_0$ and $A_1$, and similarly Bob can choose between $B_0$ and $B_1$, the observed correlators are $\{ \mean{A_xB_y}\}_{x,y}$, defined similarly to Eq.~\eqref{eq:LGI}. One would like to test whether the observed statistics comes from a local hidden variable (LHV) model, namely,
\begin{equation}\label{eq:lhv}
p(ab|xy)=\sum_\lambda p(\lambda) p(a|x,\lambda) p(b|y,\lambda),
\end{equation}
or equivalently \cite{Fine:1982PRL} there exists a global distribution $p(a_0,a_1,b_0,b_1)$ giving the $p(a_x,b_y)$ as marginals, e.g.,
\begin{equation}\label{eq:lhv_fine}
\begin{split}
p(a_0, b_0)=\sum_{a_{1} b_{1}} p(a_0, a_1, b_0, b_1).
 \end{split}
\end{equation}
If that is the case, then the correlators satisfy the CHSH inequality
\begin{equation}
\mean{A_0 B_0} +\mean{A_0B_1} +\mean{A_1B_0} - \mean{A_1B_1}\leq 2.
\end{equation}
In the temporal scenario, an analogous LGI can be derived, by considering a measurement $Q_i:=Q(t_i)$ at four time steps $t_1,\ldots,t_4$, namely
\begin{equation}\label{eq:LGI4}
\mean{Q_1 Q_2} +\mean{Q_2 Q_3} +\mean{Q_3 Q_4} - \mean{Q_1Q_4}\leq 2.
\end{equation}

Despite the analogies, the temporal and spatial scenario have a fundamental difference. While one can never perform a joint measurement of $A_0$ and $A_1$, or $B_0$ and $B_1$, as they are incompatible measurements, in the temporal scenario nothing forbids us from performing a sequential measurement of all time steps, namely, to observe $p(q_1,q_2,q_3,q_4)$. From the AoT and NSIT conditions discussed above, one can straightforwardly show that this distribution is analogous in the temporal scenario to the LHV model appearing in Eq.~\eqref{eq:lhv_fine}. We recall that Bell inequalities arise as a projection of the probability simplex, i.e., the set $\mathcal{P}=\{ \mathbf{p}\in \mathbb{R}^n\ |\ \sum_i p_i = 1, \ p_i\geq 0 \ \forall i\}$,  associated with a global distribution over all variables, e.g., via Fourier-Motzkin elimination~\cite{BudroniJPA2012}. This projection is needed to write the constraints only in terms of the statistics that can be directly observed, e.g., measurements of pairs $A_i,B_j$ in the Bell scenario. The most commonly used LGIs, such as \cref{eq:LGI,eq:LGI4}, can be defined in a similar way as the hyperplanes delimiting the projection of the simplex associated with the global distribution of measurements over all times.  However, since this global distribution, e.g., $p(q_1,q_2,q_3,q_4)$ in the above example, is directly observable, there is in principle no need to use these LGIs. 

Let us make a concrete example to clarify this point. Let us assume we are in the experimental scenario where the correlators appearing in Eq.~\eqref{eq:LGI4} are observed, i.e., $\mean{Q_i Q_j}_{\rm obs}$,  and we want to understand whether they admit a MR model. If there are no restrictions on the possible measurements we can perform, the easiest way is to measure the probability for the full sequence $p_{\rm obs}(q_1,q_2,q_3,q_4)$ and verify whether the observed correlators are compatible with it, for instance, verify that $\mean{Q_3 Q_4}_{\rm obs}=\sum_{q_1,q_2, q_3, q_4} q_3 q_4 \ p_{\rm obs}(q_1,q_2,q_3,q_4)$.

In more abstract terms, we can say that MR correlations are completely described by the probability simplex associated with the global distribution $p({\bf{q}}|1,\ldots,1)$ over all possible measurements in a sequence and all MR conditions are satisfied if and only if the statistics of shorter sequences can be interpreted as marginals of such a global distribution. This follows directly from the AoT and NSIT conditions discussed in the previous subsection. Similar conclusions on the simplicial structure of MR models are presented in Ref.~\cite{Schmid2022} by investigating the problem from a generalized-probability-theory perspective.
	
In this sense, one can claim that LGIs in the usual formulation as inequalities on some marginals of the global distributions, such as Eq.~\eqref{eq:LGI4} derived in analogy with Bell inequalities,  are not needed to detect nonmacrorealism. Moreover, correlations in a LG experiment are fundamentally different, as observed by Clemente and Kofler~\cite{ClementePRL2016}, since they do not necessarily have compatible marginals, as the NSIT condition may be violated in the first place. 

It is a matter of terminology, however, what the distinction between LGIs and NSIT conditions is. Clemente and Kofler define LGIs in analogy with Bell inequalities, in terms of the projected hyperplanes of the probability simplex, but,  arguably, this choice can be motivated only by historical reasons.
If one accepts the broader definition of Emary {\it et al.} review \cite{EmaryRPP2014}, LGIs are ``a class of inequalities [...] that any system behaving in accord with our macroscopic intuition should obey.'' This would include also NSIT conditions, which represent necessary and sufficient conditions for the existence of a MR model compatible with the observed correlations, provided that a sequence of measurements at all times considered is performed, e.g.,  $p(q_1,q_2,q_3,q_4)$ in the example above. 

We recall that the question of MR is always the following: are the observed correlations compatible with a MR model? In this sense, there may be practical limitations on the admissible measurable sequences, e.g., it may be impossible to perform more than two measurements in a row. In this case, one may still use ``standard LGIs'' in combination with NSIT conditions to conclude that the observed correlations are compatible or incompatible with a MR model.


Finally, we remark that the notions of NSIT and AoT need not be restricted to the case of a single physical quantity evolving in time, corresponding to the choice of settings $1$ (measurement) or $0$ (no measurement). For instance, the case of several different measurements at each instant of time is still consistent with the notion of macrorealism and noninvasive measurability for multiple physical quantities, as well as with the notion of  nonsignaling in time.


\subsection{Quantum correlations in time}

The most general quantum measurement is described by a positive operator valued measure (POVM), namely, a collection of operators $\M=\{M_{q}\}_q$, one for each outcome $q$, that are positive semidefinite, i.e., $M_q\geq 0$, and sum up to the identity, i.e., $\sum_q M_q = \openone$. POVMs provide a way of computing the probability of an outcome $q$ given an initial state $\varrho$ via the Born rule, i.e., $p(q)=\tr[\varrho M_q]$. 
We recall the basic properties of quantum measurements; see, e.g., \cite{HeinosaariZiman2011} for more details. Standard projective measurements are a special case of POVMs, in which each {\it effect} $M_q$ is a {\it projector}: $M_q^2= M_q:=\Pi_q$. These special POVMs are also called {\it projection valued measures} (PVMs). A state update rule is also associated with projective measurements, called the {\it L\"uders' rule} \cite{Luders:1951APL}, given by the transformation $\varrho \mapsto \Pi_q \varrho \Pi_q$, whenever the outcome $q$ is observed. Notice that the resulting state is subnormalized, and its normalization $\tr[\Pi_q \varrho \Pi_q]=\tr[\varrho \Pi_q]$ precisely represents the probability of observing the outcome $q$. The L\"uders rule can also be defined for general POVMs as the transformation $\varrho \mapsto \sqrt{M_q}\varrho \sqrt{M_q}$, whenever the outcome $q$ is observed. Again, notice that the state is subnormalized, with the normalization representing the probability of observing the outcome $q$. L\"uders rule, however, is not the most general state-transformation that follows a measurement. In general, a transformation associated with a POVM is described by a {\it quantum instrument} $\{\II_q\}_q$, where $\II_q$ is a completely positive (CP) map for all $q$ and $\sum_q \II_q$ is a CP and trace-preserving map. 

Quantum instruments, as all CP maps, can be defined via their Kraus representation, namely $\II_q(\varrho)=\sum_i K_i^q \varrho {K_i^q}^\dagger$, where $\{K_i^q\}_i$ are the Kraus operators associated with the outcome $q$, and all together they satisfy the relation $\sum_{i,q} {K_i^q}^\dagger K_i^q=\openone$. Notice how the L\"uders rule corresponds to a special choice of Kraus operators, i.e., $\{K_i^q\}_i=\{ \sqrt{M_q} \}$, but this choice is not unique. In fact, any instrument that satisfies $\sum_i {K_i^q}^\dagger K_i^q= M_q$ for all $q$ is an instrument compatible with $\M$.

All quantum measurements are, in principle, admissible for LG tests, provided that the NIM assumption can be reasonably justified. Indeed, we see in the next section that in some cases non-projective measurements, as well as state-update rules more general than the L\"uders rule provide perfectly reasonable alternatives.

\subsubsection{Temporal correlations for projective measurements}
In this section, we summarize how to compute temporal quantum correlations for the case of projective measurements, when the dimension of the system is unrestricted. This method can be used, among other things, to calculate maximal violations of LGIs under projective measurements and the L\"uders rule.
Consider a sequence of measurements $\mathbf{s}=(s_1,\ldots,s_n)$ and outcomes $\mathbf{q}=(q_1,\ldots,q_n)$, each obtained via measurements on a quantum system through a PVM $\{ \Pi_{q|s}\}_q$. The probability of the sequence can be written as
\begin{equation}\label{eq:seq_proj}
\begin{split}
p(\mathbf{q}|\mathbf{s})=\tr\left[\Pi_{q_n|s_n} \ldots \Pi_{q_1|s_1} \varrho \Pi_{q_1|s_1} \ldots \Pi_{q_n|s_n} \right] \\
= \tr\left[ \Pi_{\mathbf{q}|\mathbf{s}}^\dagger \varrho \Pi_{\mathbf{q}|\mathbf{s}} \right] = \tr\left[  \varrho \Pi_{\mathbf{q}|\mathbf{s}} \Pi_{\mathbf{q}|\mathbf{s}}^\dagger \right],
\end{split}
\end{equation}
where we defined $\Pi_{\mathbf{q}|\mathbf{s}}:=\Pi_{q_1|s_1} \ldots \Pi_{q_n|s_n}$. Coming from a PVM, the operators $\{ \Pi_{q|s} \}_s$, for any given $s$, satisfy additional constraints, namely, $\Pi_{q|s}=\Pi_{q|s}^\dagger$ for all $q,s$, $\Pi_{q|s}\Pi_{q'|s}=\delta_{qq'}\Pi_{q|s}$ and $\sum_{q} \Pi_{q|s}=\openone$, i.e., hermiticity, orthogonality, idempotence, and completeness.

Whenever we have a quantum state $\varrho$ and a collection of PVMs $\{\Pi_{q|s}\}_{q,s}$ we can define the matrix
\begin{equation}\label{eq:MM_seq}
C_{\bq|\bs;\bqp|\bsp} = \mean{\Pi_{\mathbf{q}|\mathbf{s}} \Pi_{\mathbf{q'}|\mathbf{s'}}^\dagger},
\end{equation}
where the expectation value is taken with respect to the state $\varrho$, i.e., $\mean{\cdot}=\tr[\cdot \varrho]$. Such a matrix satisfies two constraints: (a) $C\geq 0$, i.e., it is positive semidefinite, since $v^\dagger C v$ can be written in the form $\mean{B^\dagger B}$ for any vector $v$, (b) $C$ satisfies a number of linear constraints coming from the linear constraints of the PVMs, i.e., hermiticity, orthogonality, idempotence, and completeness, meaning that certain elements will either be equal to each other or equal to zero. As we showed in Eq.~\eqref{eq:seq_proj}, diagonal elements correspond to observable probabilities.

We call $C$ a {\it moment matrix} associated with outcomes $\{q\}_q$, settings $\{s\}_s$ and length $L$ if it is a positive semidefinite matrix that satisfies the linear constraints discussed above, where all elements associated with probabilities of sequences of measurements of length $L$ appear, i.e., $C_{\bq|\bs;\bq|\bs}$, for all $\bq$ and $\bs$ vectors of length $L$.  This implies that any optimization of a linear function of these probabilities, such as the maximal quantum violation of an LGI via projective measurements can be upper bounded via semidefinite programming (SDP)~\cite{Boyd2004Convex}, based on the {\it moment matrix} defined by Eq.~\eqref{eq:MM_seq}. However, a stronger result holds, namely, that for any moment matrix satisfying the positivity and linear constraints discussed above, one can reconstruct the quantum state and PVMs that provide the diagonal matrix entries as probabilities. 
This implies that the bound obtained with this method is exact. 

It is important to remark that this method makes no assumption on the dimension of the quantum system. In fact, an explicit solution, i.e., quantum state and PVMs, can be extracted from the explicit solution of the SDP, i.e., an optimal matrix $C^*$,  and it corresponds to a Hilbert space dimension equal to the rank of $C^*$. This is the method for bounding temporal correlations for projective measurements introduced in \cite{BudroniPRL2013}. 

A typical example application of this method is to find the quantum bound of $3/2$ for the three-term LGI of Eq.~\eqref{eq:LGI} and, more generally, for the $N$-term inequality
\begin{equation}
\sum_{i=0}^{N-1} \mean{Q_i Q_{i+1}} - \mean{Q_0 Q_{N-1}} \stackrel{\rm MR}{\leq} N-2\stackrel{\rm QM}{\leq} N\cos\left(\frac{\pi}{N}\right), 
\end{equation}
which can be derived analytically, see \cite{Araujo:2013PRA} for the classical and \cite{BudroniPRL2013} for the quantum bound.

These quantum bounds are derived under the assumption of projective measurements and the L\"uders rule. However, if more general measurements and transformation rules (i.e., quantum instruments) are considered and no additional constraints are imposed, trivial bounds appear. This happens also in the case of projective measurements, if a different transformation rule, called the {\it von Neumann rule} is used. According to von Neumann's original formulation of the projection postulate \cite{vonNeumann:1932SPR}, the state is always updated through rank-1 projections, even if the projector associated with a given outcome (i.e., eigenvalue of the physical observable) is degenerate. This rule was firmly criticized by L\"uders~\cite{Luders:1951APL}, who introduced what is now the textbook projection rule. Nevertheless, the von Neumann state-update rule can still be a valid description of a physical situation, for instance, when the system strongly interacts with the measurement apparatus, but we are not able to properly read the classical outcome. Think about a Stern-Gerlach measurement of a spin-$j$ particle in which we are only able to assess whether the particle hit the upper or the lower part of the screen. This corresponds to a measurement on $N=2j+1$ quantum levels, which are then coarse-grained at the level of classical outcomes into just  two. 
In this case, different bounds can appear, which now explicitly depend on the system dimension $N$. In fact, the higher the dimension, the higher the number of outcomes that can be coarse grained. 

Optimal values for the corresponding LGI, or any other linear function of the probabilities, can be computed for a given dimension of the Hilbert space with a combination of upper bound (the SDP method discussed above) and explicit numerical or analytical solutions~\cite{BudroniPRL2014}. 
In particular, it was shown~\cite{BudroniPRL2014} that for a spin-$j$ particle the following value is achieved for the three-term LGI of Eq.~\eqref{eq:LGI}:
\begin{equation}
\mean{Q_1 Q_2} + \mean{Q_2 Q_3} -\mean{Q_1 Q_3} = 3-\sqrt{\frac{2}{\pi j}},
\end{equation}
which tends to the algebraic bound of $3$ for the limit 
$j\rightarrow \infty$, and where the measurement is coarse-grained according to the relabeling $-1$ for the outcome $-j$ and $+1$ for all other outcomes.

A very elegant treatment of the same phenomenon, providing exact and analytical bounds for any dimension, is given by Schild and Emary~\cite{SchildPRA2015} for the case of the quantum witness~\cite{LiSR2012} or equivalently the NSIT condition for a sequence of length 2~\cite{KoflerPRA2013}, namely,
\begin{equation}
W = | p_2(q_2) - \sum_{q_1} p_{12}(q_1 , q_2) |
\end{equation}
for a fixed outcome $q_2$. The witness $W$ ranges from $0$, for MR models, to $1$, the algebraic maximum imposed by the fact that $p_2$ and $p_{12}$ are probabilities. Schild and Emary~\cite{SchildPRA2015} showed that for a $N$-level quantum system subject to projective measurements with the von Neumann state-update rule, one can obtain
\begin{equation}
W_{\max}^{\rm vN} = 1-\frac{1}{N},
\end{equation}
which, again, tends to the algebraic maximum in the limit of infinite dimension. Finally, a systematic treatment of this phenomenon for different types of classical post-processing, and including a discussion on possible experimental realizations, has been provided by Lambert {\it et al.}~\cite{LambertPRA2016}.

From this observation, one can already see that the minimal dimension needed for the description of a quantum experiment can be learned from the value of an LGI expression. Another similar
approach to the certification of the dimension of a quantum system, which combines the moment matrix approach of Budroni and Emary~\cite{BudroniPRL2014} together with the Navascu\'es-Vertesi method for imposing dimension constraints on moment-matrix approaches, was presented by Sohbi {\it et al.}~\cite{Sohbi2021}. In contrast with the approaches previously presented~\cite{BudroniPRL2014,SchildPRA2015,LambertPRA2016}, this paper does not involve any classical coarse graining of the measurement outcomes.
We also discuss more concretely applications of temporal correlations to witness the dimension of a physical system afterwards, in \cref{sec:applications}.

\subsubsection{Temporal correlations for more general measurements}\label{sssec:TC_gen}
The situation changes drastically for the case of nonprojective measurements, more precisely, for measurements that do not obey the L\"uders (or von Neumann) state-update rule. In fact, Fritz~\cite{FritzNJP2010} showed that any correlation belonging to the AoT polytope can be achieved by quantum systems if enough internal memory is available; see also the discussion in \cite{ClementePRL2016}.

Since this result holds in more general settings than usual LG tests, it is convenient to take a step back and define the AoT polytope for arbitrary measurement inputs, instead of just $0$ (no measurement) and $1$ (measurement), as discussed so far. In order to understand the general idea, it is sufficient to look at the simple case of sequences of length 2 described by a distribution $p(ab|xy)$. In particular, we follow the presentation in \cite{Hoffmann2018}. Due to the AoT constraints, we have that
\begin{equation}
\sum_b p(ab|xy)=\sum_b p(ab|xy'), \text{ for all } y,y'.
\end{equation}
This implies that we can always define $p(a|x):=\sum_{b} p(ab|xy)$ and $p(b|a;xy):= p(ab|xy)/p(a|x)$, defined to be zero if $p(a|x)=0$. We can then write
\begin{equation}\label{eq:aot_2times}
p(ab|xy)= p(a|x)p(b|a;xy) .
\end{equation}

From this expression, it is clear that any pair of deterministic strategies, i.e., distributions $\{p(a|x)\}_x$, $\{p(b|a;xy)\}_{a,x,y}$ with values $0$ or $1$, correspond to extreme points of the AoT polytope, as they cannot be further decomposed. Conversely, arbitrary nondeterministic distributions $\{p(a|x)\}_x$, $\{p(b|a;xy)\}_{a,x,y}$ can always be decomposed as convex mixtures of deterministic ones. This implies that the extreme points of the AoT polytope are all and only the deterministic strategies \cite{HoffmannThesis2016,Hoffmann2018,AbbottPRA2016}. In particular, this also implies that all temporal correlations can be reproduced by classical systems, in stark contrast to the spatial case. 

The first observation is that such strategies clearly involve an invasive measurement: The outcome-generation strategy at later time steps explicitly depends on the previous time steps. The second is that the resource necessary for generating such correlations is given by the number of internal states of the system: 
One needs to store the information about the past inputs and outputs, e.g., $(a,x)$ in the example in Eq.~\eqref{eq:aot_2times}, in different states in order to generate the correct output at the subsequent time-steps. Moreover, it is also clear that any classical model can be simulated by a quantum one, for instance, with projective measurements in a fixed basis followed by unitary rotations corresponding to the classical state transition.  

These results can be used to investigate temporal correlations beyond the original approach of Leggett and Garg by relaxing the assumption of noninvasive measurements and substituting it with the condition of finite  number of states. The NIM condition, then, simply corresponds to the case of a single internal state (i.e., zero internal memory). A detailed discussion of this approach is presented in \cref{sec:finitestates}.

\subsubsection{Temporal steering}\label{sec:Tsteeering}
A different approach to temporal correlations that combines elements of macrorealist and quantum models is that of {\it temporal steering}. Chen et al.~\cite{ChenPRA2014} introduced the concept of temporal steering inspired by the analogous notion in the spatial scenario \cite{UolaRMP2020}. Instead of looking at a macrorealist model for the observed correlations, as in standard LG tests, one assumes a higher control over the physical system that is necessary to perform state tomography. The object of a temporal steering test is then the {\it state assemblage} $\{\sigma_{a|x}\}_{a,x}$ obtained by performing measurements described by instruments $\{\II_{a|x}\}_{a,x}$. More precisely, for an initial state $\varrho$ we have
\begin{equation}
\sigma_{a|x} := \II_{a|x}(\varrho).
\end{equation}
Notice that $\sigma_{a|x}$ is subnormalized, i.e., $\tr[\sigma_{a|x}]\leq 1$, where the normalization factor represents the probability of the outcome $a$.
The question of temporal steering is then whether the assemblage $\{\sigma_{a|x}\}_{a,x}$ admits a {\it temporal hidden state model}, namely,
\begin{equation}\label{eq:THS}
\sigma_{a|x} = \sum_\lambda p(\lambda) p(a|x,\lambda) \tilde{\sigma}_\lambda,
\end{equation}
where $\{\tilde{\sigma}_\lambda\}_\lambda$ is a set of normalized states and in which case the assemblage is  {\it unsteerable}. In other words, a state assemblage is unsteerable if it can be interpreted as coming from a classical postprocessing, given by the distribution $p(a|x,\lambda)$, of an initial collection of states $\{\tilde{\sigma}_\lambda\}_\lambda$ distributed according to $p(\lambda)$. It is clear that an unsteerable assemblage satisfies all LG conditions. In fact, for an assemblage described by \cref{eq:THS} it is impossible to detect which measurement has been performed, as the state resulting from discarding the outcome is always $\sum_\lambda \tilde{\sigma}_\lambda$, regardless of the input $x$.

Temporal steering needs a stronger set of assumptions and a more detailed characterization of the physical system since we need to assign a quantum state to each measurement outcome. At the same time, this allows the use of temporal steering for a broader range of quantum information applications. For instance, temporal steering has been applied to the analysis of the security bounds in quantum cryptography \cite{ChenPRA2014, BartkiewiczPRA2016}, the quantification of non-Markovianity~\cite{ChenPRL2016} and causality~\cite{KuPRA2018}, and the investigation of the radical-pair model of magnetoreception~\cite{KuPRA2016}. Finally, the relation between  spatial, temporal, and channel steering has been investigated in \cite{UolaPRA2018}, where also a unified mathematical framework has been introduced; see also \cite{Pusey2015,Ku2022}.

\section{Experimental tests of macrorealism}\label{Sec:exp_tests}
A successful experimental violation of macrorealism should consist in a clear demonstration of a preparation of a superposition between two ``macroscopically distinct'' states. For example, a large massive object in a superposition of two (macroscopically) different locations. 
However, several practical difficulties arise even at the stage of designing such a test, from the very definition of macroscopically distinct states, to the issue of performing measurements that are convincingly nondisturbing from a macrorealist perspective. In this section, we review the approaches designed to overcome typical loopholes, prominently the so-called clumsiness loophole~\cite{WildeMizel2012}, together with the experiments that have been performed in recent years.

\subsection{Loopholes in macrorealist tests}\label{sec:loopholes}

As in the case of Bell tests, practical tests of macrorealism suffer from loopholes. For example in a Bell test, a problem may arise when the detector does not register all events, which imposes, under a certain efficiency threshold, to either  assume that the registered events are a {\it fair sampling} of all possible events or admit the possibility of a local hidden variable description~\cite{PearlePRD1970,LarssonJPA2014}. Analogously, a low detection efficiency in a LG test may open the door for MR explanations that take advantage of the undetected events.

\begin{figure}
\includegraphics[width=\linewidth]{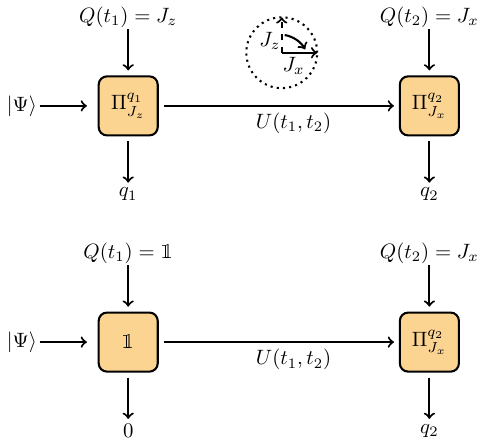}
	\caption{Two simple sequences of two measurements that lead to a violation of macrorealism: From an initial state $\ket{\Psi}$ either a projective measurement of $J_z$ or no measurement is performed, followed by a unitary rotation $J_z \mapsto J_x$ and a projective measurement of $J_x$. For an initial superposition state $\ket{\Psi}=\tfrac 1 {\sqrt 2}(\ket{1}_z+\ket{-1}_z)$ violation of macrorealism can be observed.}\label{fig:simple_example}
\end{figure}

Another important loophole in Bell experiments is the {\it locality loophole} appearing when the choice of the measurement setting on  one particle is not space-like separated from the measurement performed on a distant one~\cite{AspectPRL1982,LarssonJPA2014}. This allows for an explanation in terms of communication or influences traveling at the speed of light or below.
The corresponding situation in the temporal scenario is the violation of the NIM assumption. Here, we no longer have spatial separation among the different measurements, so without any additional assumption, one may admit the possibility that past measurements influences future ones. 
To make matters worse such influences do not even need to be conspiratorial to violate some LG condition. Simply performing a clumsy (i.e., noisy) measurement is enough to obtain a violation of LGIs. 
In fact, all temporal correlations arising from a quantum model can be simulated with a classical hidden variable theory that allows for measurements disturbance. We already discussed this at an abstract level in Sec.~\ref{sssec:TC_gen}. Let us see this now with a paradigmatic example; see \cref{fig:simple_example}.
Consider a sequence of two measurements of a (ideally macroscopic) quantum observable at two time steps, i.e., $\Q(t_1)=\{M_{t_1}^{q}\}_q$ and $\Q(t_2)=\{M_{t_2}^{q}\}_q$, on a quantum state $\varrho$.
According to the Born rule we can compute the joint probability of outcomes $(q_1,q_2)$ as
\begin{equation}
p(q_1 , q_2) = \Tr\left[M^{q_2}_{t_2} \mathcal{I}^{q_1}_{t_1}(\varrho)\right] ,
\end{equation}
where we indicated with $\mathcal{I}^{q_1}_{t_1}(\cdot)$ the quantum instrument associated with outcome $q_1$ of observable $\Q(t_1)$. On the other hand, when the first measurement is not performed we have 
\begin{equation}
p(q_2) = \Tr\left[M^{q_2}_{t_2} \varrho \right] .
\end{equation}
As we mentioned, macrorealism is equivalent to NSIT, i.e., it implies the condition
\begin{equation}\label{eq:NSIT12}
p(q_2)=\sum_{q_1} p(q_1 , q_2). 
\end{equation}
A violation of \cref{eq:NSIT12} can be observed when the triple $(\{M^{q_2}_{t_2}\}_{q_2}, \{\mathcal{I}^{q_1}_{t_1}\}_{q_1}, \varrho)$ is such that $\Tr\left[M^{q_2}_{t_2} \varrho \right] \neq \sum_{q_1} \Tr\left[M^{q_2}_{t_2} \mathcal{I}^{q_1}_{t_1}(\varrho) \right]$.
This can be interpreted as caused by any of the three elements. In particular, it can be realized even in a trivial scenario where $\Q(t_1)$ and $\Q(t_2)$ are compatible (e.g., projective commuting) observables and $\varrho$ is one of their (pure) eigenstates, with just the instrument $\{\mathcal{I}^{q_1}_{t_1}\}_{q_1}$ being clumsy (i.e., noisy).
On the other hand, the effect that we would like to witness should be ideally due to a preparation that is made on a state that is a superposition of two (macroscopically) distinct eigenvalues of $\Q$ and we should ensure that the measurement just reads off the value of $\Q$ without disturbance. 
With this, we emphasize again that the notion of macrorealism is deeply connected with the preparations as well as measurement implementations. Even quantitatively, these different effects cannot be distinguished by witnesses of macrorealism alone. Therefore, since the introduction of the LGIs, different ideas for proving that a measurement is nondisturbing from a macrorealism perspective have been discussed.

\subsubsection{Ideal-negative-result measurements}
The first idea, due to Leggett and Garg themselves~\cite{LeggettPRL1985}, was to use  {\it ideal-negative-result measurements}: The measurement apparatus is made such that it interacts with the system only when one of two outputs (say, $+1$) is observed, e.g., a microscope that detects a particle in a precise position only, which is afterwards discarded, while only null results are kept, implying that the opposite output (say, $-1$) has been observed without interacting with the system.  
A macrorealist would then believe that the value $-1$ was preexistent to the measurement, e.g., in a two-well potential, a particle not detected in one well must have been in the other.
However, a number of criticisms to this approach can be made (see, e.g., \cite{EmaryRPP2014}), and can be traced back even to early discussions about quantum measurements~\cite{Dicke81}. In short, in this approach one still has to make a rather strong assumption on the interaction model behind the measurement process. For example, a particle that is not visible at the microscope may have just absorbed the photons.
This approach has been employed in several experiments, from the early ones~\cite{EmaryRPP2014} to the recent ones performed with single particles making random walks on an optical lattice~\cite{RobensPRX2015}, in nuclear spins~\cite{KatiyarNJP2017,MajidyPRA2019} and in heralded photons~\cite{JoarderPRX2022}. In some cases~\cite{MajidyPRA2019}, the negative results measurements were employed as a benchmark to support the nondisturbance arguments made with other approaches.

\subsubsection{Weak and ambiguous measurements}
{\it Weak measurements}  
arise from models with a small interaction between system and measurement apparatus.
A typical concrete model for such a detector is obtained by means of an ancillary system usually considered as continuous variable with quadrature $x$ in a Gaussian state $|\phi(x)\rangle=\int{\rm d}x \  \phi(x) |x\rangle $, with $\phi(x)=(2\pi s)^{-1/4}\exp(-x^2/4s^2)$ being a Gaussian wave function with standard deviation $s$. The canonical form of a POVM associated with the weak measurement of a projective observable $\Q$, i.e., an Hermitian operator with spectral decomposition $\Q=\sum_q q\ketbra{q}$, is then given by \cite{WV_Review,EmaryRPP2014} $\W(\Q) = \{W_x \}_x= \{ K_x^\dagger K_x \}_x$, with Kraus operators
\begin{equation}\label{eq:Kq_def}
K_x = (2\pi \sigma)^{-1/4}\exp(-(x-\Q)^2/4s^2).
\end{equation}
Here, $s$ gives the weakness of the measurement: In the limit $s \rightarrow \infty$ one obtains a weak measurement, while in the limit $s \rightarrow 0$ one obtains the projective measurement $\Q$ itself. Note also that the outcomes $x$ are referred to the ancillary system and are different from $q$. In fact, $x$ is usually a continuous variable whereas $q$ can be from a discrete set of outcomes.
Still, the Kraus operators can be defined in a way such that the expectation value of $\W(\Q)$ coincides with that of $\Q$, i.e., $\int x \Tr[W_x \varrho] {\rm d} x = \sum_q q \Tr[\Pi_q \varrho]$.

With this idea in mind, continuous weak measurements have been considered in early experiments~\cite{PalaciosNatPhot2010,EmaryRPP2014} and in more recent approaches~\cite{HalliwellPRAContMeas, MajidyPRA2019}. The continuous version of weak measurements can be modeled by a linear input/output relation between the measured variable and the inferred one
\begin{equation}\label{eq:W_weak}
	W(t) = a \Q(t) + b(t) ,
\end{equation}
where $b(t)$ is a noise term that is typically assumed uncorrelated with $\Q(t)$, which corresponds to a vanishing back-action on the system in the limit of a weak measurement, i.e., $a \rightarrow 0$. 
For instance, in Refs.~\cite{HalliwellPRA2016,HalliwellPRADecoherent} the quasiprobability approach is combined with weak measurements and ideal-negative results to justify the NIM assumption in an experimental setting. 

Note however, that here NIM still relies mostly on the weakness of the measurement in itself, and
in fact, in some experiments it has been employed complemented with the negative-result measurement idea~\cite{MajidyPRA2019}. 
In this respect, note once more that a quantum model for a weak measurement has no meaning for a stubborn macrorealist.  In a macrorealist perspective the measurement process is described in a device-independent fashion based solely on MRPS, NIM, and IND assumptions. 

To address this problem, Emary~\cite{EmaryPRA2017} proposed to substitute the NIM assumption with a different one, namely the {\it equivalently invasive measurability} (EIM) assumption:
\begin{itemize}
\item[EIM] The invasive influence of ambiguous measurements on any given macroreal state is the same as that of unambiguous ones
\end{itemize}
Here, an {\it ambiguous} measurement is intended as one not revealing the exact value of an observable, but rather giving some noisy information. One example is given by Eq.~\eqref{eq:W_weak}, where the term $b(t)$ is a noise term uncorrelated with the actual measurement of the observable $\Q(t)$. 
The simplest example of an ambiguous measurement discussed by Emary~\cite{EmaryPRA2017} is the following. Instead of performing a projective measurement $\Q = \sum_{q=0}^2 q\ketbra{q}$, one considers the three projectors $\{\openone - \ketbra{q}\}_{q=0}^2$. Each of them detects the state in which the system is not, which we denote as $\overline{q}$, with a certain probability $p_A(\overline{q})=\tr[\varrho (\openone -\ketbra{q})]$. Note that $\{\openone - \ketbra{q}\}_{q=0,1,2}$ is not a valid POVM, thus the measurement process involves some postselection of the favorable outcomes. 

From a macrorealist perspective, identifying $\ket{0},\ket{1}$, and $\ket{2}$ with the possible states of the system, one can recover the probability as
\begin{equation}\label{eq:p_A}
p(q) = \tfrac{1}{2} \left[p_A(\overline{q}') + p_A(\overline{q}'') - p_A(\overline{q})\right],
\end{equation}   
for every triple of distinct values $q,q',q''=0,1,2$. Note that this is, in principle, a quasiprobability distribution, i.e., it may become negative depending on how these quantities are measured. However in quantum mechanics for the specific choice of $\Q$ above this identification works at the level of probabilities, namely
\begin{equation}
\begin{split}
&\tfrac{1}{2} \left( \openone - \ketbra{q'} + \openone-\ketbra{q''} -\openone + \ketbra{q}\right)\\
&=\tfrac{1}{2} \left( \openone - \ketbra{q'} -\ketbra{q''} + \ketbra{q}\right) = \ketbra{q}.
\end{split}
\end{equation}

EIM then assumes that the disturbance caused by this measurement is the same as the one caused by the direct measurement of $\Q$. 
In practice, one performs first the ambiguous measurement, with the L\"uders instrument, and then the projective one. Then, via the relation in Eq.~\eqref{eq:p_A} it is possible to estimate the probability $p(q_1 , q_2)$.
EIM assumes that the disturbance from the direct measurement, defined via the NSIT condition
\begin{equation}
\delta(q_2) = p(q_2) - \sum_{q_1} p(q_1 , q_2),
\end{equation} 
is the same as the one from the ambiguous measurement
\begin{equation}
\delta_A(q_2) = p(q_2) - \sum_{q_1} p_A(q_1 , q_2).
\end{equation} 
This quantity becomes then the correction to the three-term LGI, with measurements $\Q(t_i)$ for $i=0,1,2$ and the first measurement being trivial, namely
\begin{equation}\label{eq:LG_EIM}
\mean{Q_1}+\mean{Q_1 Q_2}-\mean{Q_2} \leq 1 + \sum_{q_2} |\delta_A(q_2)|.
\end{equation}
This expression is intended to be obtained by estimating the quantity $\delta_A(q_2)$ via the ambiguous measurement and the correlators $\mean{Q_i}$ and $\mean{Q_1Q_2}$  via the unambiguous one.
It was shown theoretically~\cite{EmaryPRA2017} and experimentally~\cite{WangPRA2018} that this inequality can be violated, thus proving a violation of macrorealism based on the EIM, rather than NIM.
The same approach can be extended to include weak measurements, but not to systems of lower dimension: at least a three-level system is needed \cite{EmaryPRA2017}.

\subsubsection{Quantum nondemolition measurements}
As we mentioned, all of the proposals based on weak measurements rely on the idea that in such models the system is weakly perturbed after the measurement. We recall that an ideal weak measurement should be understood in the limit, e.g., $s \rightarrow \infty$ in the model \eqref{eq:Kq_def}. 
However, it has been argued \cite{EmaryRPP2014,UolaPRA2019} that a weak measurement by itself does not guarantee nondisturbance. In contrast, even a strong measurement, i.e., projective, may be nondisturbing with respect to the statistics obtained by certain subsequent measurements. This is the case, for instance, of a sequential measurement of commuting projective observables.
An example of this is given by the {\it quantum nondemolition measurement} (QND)~\cite{braginsky}.
In general terms, a QND measurement is performed via a probe that interacts with the system in a way that the measured variable is not disturbed, e.g., via an interaction Hamiltonian $H_I$ 
such that $[H_I,\Q]=0$. 
The idea is that even in the presence of noise, the observable $\Q$ should not be perturbed by the measurement, as it can be shown via rapid and repeated measurements of it~\cite{Sewell2013}. In this sense, the QND measurement represents a practical implementation of the ideal projective measurement in a nondestructive way, which then allows one to repeat the measurement multiple times. This makes it ideal for LG tests.  In fact, this direction has been explored in early works on macrorealism~\cite{CALARCO1995279,CalarcoAPB1997}, which reached the conclusion that LGIs violations could not be observed with QND measurements. Subsequent works, however, showed that QND measurements can indeed lead to a violation of LGIs, even in macroscopic systems~\cite{LG_QNDPRL2015,RosalesZaratePRA18}. 

From a quantum-mechanical perspective what happens is that, even if $\Q$ is not perturbed, some other incompatible variable (let us denote it by $\P$) is perturbed. A paradigmatic example is the QND measurement of a particle's position $\Q$, which disturbs its momentum $\P$, as in the original Heisenberg's microscope model~\cite{Heisenberg:1927ZP}. This is, of course, not enough to conclude that the measurement is completely noninvasive. Nevertheless, confronted with these facts, a macrorealist would conclude that the disturbance is not on $\Q$, but in some other variable, $\P$, which then transforms into $\Q$ due to the time evolution. In other words, to explain the experiment a macrorealist would be forced to introduce a classical version of the quantum back-action~\cite{LG_QNDPRL2015}.

\subsubsection{Wilde-Mizel adroit measurements}\label{sssec:WM_adroit} 
Wilde and Mizel ~\cite{WildeMizel2012} introduced the terminology ``clumsiness loophole'' and proposed a possible solution via the so-called adroit measurements.
The idea is to decompose the measurement into a number of intermediate steps $\S_i$ that taken individually do not have any disturbance effect on the sequence. 
Control experiments are then made with each of the individual operations forming the full measurement and the observations should confirm their individual nondisturbance.
Thus, it is the joint effect of all operations that {\it collude} to disturb the initial state of the system. A simple example of this idea is to make a projective measurement
with the L\"uders instrument $\mathcal{I}^{q_1}_{t_1}(X)=\Pi^{q_1}_{t_1}X\Pi^{q_1}_{t_1}$ in such a way that 
$\Pi^{q_1}_{t_1}$ is decomposed into two measurements $\Pi_{s_2}^{(1)}$ and  $\Pi_{s_1}^{(1)}$, i.e.,  $\Pi^{q_1}_{t_1} = \Pi_{s_2}^{(2)} \Pi_{s_1}^{(1)}$.  One chooses $\Pi_{s_1}^{(1)}$ that commutes with the preparation $\varrho$ and $\Pi_{s_2}^{(2)}$ that commutes with the POVM elements $M^{q_2}_{t_2}$. This way, a sequence with each of the steps $\S_i$ taken individually satisfies NSIT, i.e., $\Tr\left[M^{q_2}_{t_2} \varrho \right] = \sum_{s_i} \Tr\left[M^{q_2}_{t_2} \Pi_{s_i}^{(i)} \varrho \Pi_{s_i}^{(i)}\right]$ for $i=1,2$. Such measurements are called adroit measurements \cite{WildeMizel2012}.
However, when performed together, the two processes collude to give rise to the disturbing measurement of $\Q(t_1)$. This way, the assumption of NIM is replaced with that of noncolluding measurements. Moreover, this notion can be generalized to allow for a small amount of noise, introducing the idea of $\varepsilon$-adroit measurements. Via extra control experiments, this deviation $\varepsilon$ from the ideal case can be quantified and, under certain assumptions, used to modify the macrorealist bound for the corresponding LGI.
Recently, this approach has been employed to disprove macrorealism in a superconducting qubit in the IBM cloud platform~\cite{HuffmanPRA2017}.

This idea was generalized in \cite{UolaPRA2019}, with the goal of exploring the connections between macrorealism and nondisturbance of quantum measurements, defined in a state-independent fashion and beyond the simplest case of projective measurements.
In fact, in the case of projective measurements it is known that non-disturbance coincides with commutativity of the two observable operators, i.e., that two projective measurements $\P$ and $\Q$ can be performed in a sequence that is nondisturbing regardless on the preparation, if and only if $[\P,\Q]=0$, which implies also the commutativity of all spectral projectors.
Moreover, the instrument that achieves a nondisturbing sequence is simply given by the L\"uders rule $\mathcal I_{p}^{\P}(X) =  \Pi_p X \Pi_p$. 

In other words, if two measurements are given by commuting projectors, then the L\"uders instrument gives a nondisturbing implementation of the sequence for every possible initial state, while if they are given by noncommuting projectors there is always one preparation such that some disturbance is observed, independently of which instrument is implemented. 
Moreover, sequences involving more projective measurements are nondisturbing if and only if the measurements are pairwise nondisturbing.
Note also that in this context weak measurements can be seen as classical postprocessings of projective quantum measurements, i.e, noisy versions of them, which cannot improve their nondisturbance. 
In contrast, if anything, such postprocessing might increase the disturbance; see the detailed discussion in~\cite{UolaPRA2019}.

For more general POVMs, in contrast, it is actually possible to find triples of measurements that are pairwise nondisturbing (i.e., when performed in a sequence of two of them, there exists an instrument such that NSIT is satisfied for all possible state preparations), while a sequence of all three of them would violate NSIT for some preparations. In this sense, such a triple can be seen as a state-independent generalization of the notion of adroit measurements~\cite{UolaPRA2019}. Moreover, it is conjectured in \cite{UolaPRA2019} that, in analogy with joint measurability structures, this structure of state-independent nondisturbance relations generalizes to more complex ones. This suggests that it should be possible to perform LG tests with $n$-adroit measurements, namely, measurements that are nondisturbing for any quantum state whenever a sequence of $n$ of them is measured, but they are nevertheless able to violate a LGI if sequences of $n+1$ of them are considered. Of course, this does not close completely the door on a classical explanation, but it puts much stronger constraints to the possible MR models able to describe such an experiment.

\subsubsection{Macroscopic limit: coarse-grained measurements and quantum-to-classical transition}

A crucial difference between LG and Bell tests is the emphasis of the former on macroscopic systems, namely, on the fact that quantum effects should be, in principle, observable at a macroscopic level. Nevertheless, many different notions of macroscopicity have been given in the literature, with no common agreement on a single one~\cite{MacroReview}.
Ideally, LG tests should involve observables that are ``extensive'', i.e., whose value should scale as $N$ for an $N$-body system, with a state preparation that: (i) is a superposition of two extensively different values, like $\Q=N$ and $\Q=-N$, and (ii) has large $N$.

Following this idea, the first quantifier of macroscopicity was introduced by Leggett and Garg~\cite{LeggettPRL1985}. It consists of a pair: the {\it extensive difference} that is the difference between the maximum and minimum observed values of $\Q$, divided by some relevant unit scale for the system, and the {\it disconnectivity} that loosely speaking quantifies the quantumness of the state, via, e.g., a suitable entanglement measure between the $N$ particles.

However, further elaborations of this idea have led to many different measures that are completely independent and in some cases unrelated to the LG approach, such as cat states in quantum optics; see the recent review \cite{MacroReview}.
For example, one other relevant measure that has been discussed in the context of LG tests is based on the idea of disproving certain collapse theories based on gravitational forces: a macroscopic quantum state in that sense has a large total mass and is displaced along large distances~\cite{Nimmrichter13}.
This has been discussed also in recent experiments~\cite{RobensPRX2015}.

The notion that one wants to capture is similar to the old Schr\"odinger cat idea: that microscopic quantum effects, such as superpositions, could be amplified to the macroscopic regime, as opposed to a so-called quantum-to-classical transition that should happen at a fundamental scale, as argued in collapse models~\cite{Bassietalreview13}.  In short, the question that is source of quite intense debate even to this date is whether the difficulty in detecting quantum effects at macroscopic scales is due to technological reasons, e.g., decoherence effects that increase at the macroscopic scales~\cite{SCHLOSSHAUER20191}, or to a more fundamental reason, such as a collapse mechanism~\cite{Bassietalreview13}. 
An intermediate position was given in \cite{Kofler2007}, where it was hypothesized 
that at macroscopic scales measurements must be necessarily coarse grained and that would prevent us from witnessing quantum effects. See also \cite{KimPRL2014} for a similar conclusion.
This latter idea, however, seems to be contradicted by recent works~\cite{BudroniPRL2014,LG_QNDPRL2015,SchildPRA2015,LambertPRA2016, WangPRA2016}, which showed that coarse-grained measurements do not necessarily shade away quantum effects, but in fact can even enhance their visibility.

\subsection{Recent experiments}

Macrorealism tests have been performed in recent years in a wide variety of systems. Despite them being still performed mostly on microscopic systems, we have seen in recent years a growing interest in performing experiments that take seriously the different loopholes associated with LG tests, especially the problem of NIM, and devising ingenious ways to address them. In the following, we review some of the most recent experiments, 
which were not covered by the previous review on the topic~\cite{EmaryRPP2014}.

\subsubsection{Single atoms in optical lattices}
\begin{figure}[t]
	\includegraphics[width=\linewidth]{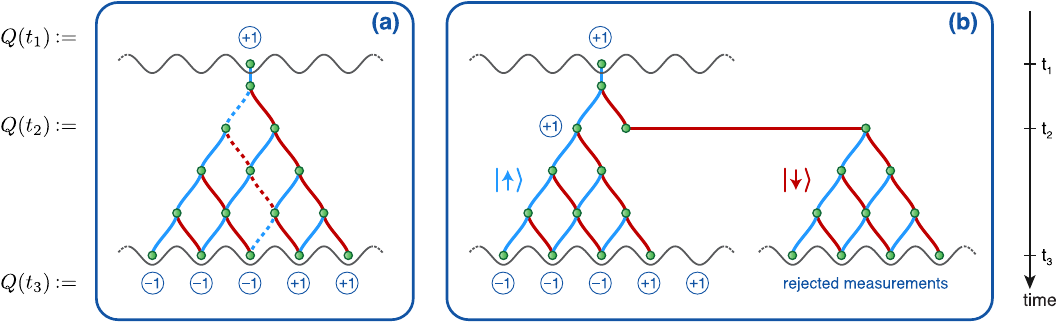}
	\caption{(Reproduced from Ref.~\cite{RobensPRX2015}). (a) A Cs atom performs a four-step random walk on a $1$D lattice with spin-dependent shifts. (b) The intermediate measurement at $t_2$ is an ideal-negative-result measurement: if the atom is spin down it interacts with the field, it is displaced far away and the measurement is discarded, otherwise it experiences no interaction.
     }\label{fig:Robens}
\end{figure}

Robens {\it et al.}~\cite{RobensPRX2015} performed a LG test with a Cs atom  in an optical lattice that performs a quantum walk; see \cref{fig:Robens}.
The experiment worked as follows: a Cs atom with two addressable internal states, $\ket{\!\!\uparrow}$ and $\ket{\!\!\downarrow}$, is controlled via two independent optical lattices. The atom ``experiences'' only the field of one of the lattices, depending on whether it is in the state $\ket{\!\uparrow}$ or in the state $\ket{\!\downarrow}$. 

Therefore, a coherent superposition $(\ket{\!\!\uparrow} + \ket{\!\!\downarrow})/\sqrt{2}$ follows a quantum random walk, i.e., two independent paths simultaneously, according to a quantum description, whereas it follows a random walk according to a MR description.
The atom is then allowed to displace four times, and measurements are performed at $t_1=\tau$, $t_2=2\tau$ and $t_3=4\tau$, where $\tau$ is approximately $26\mu s$; see \cref{fig:Robens}(a).
The LG test works as follows. At step $1$ the atom is just prepared in state $\ket{\!\!\uparrow}$ and at position $x=0$, with a deterministic assignment $\Q(t_1)=1$. 
At the intermediate time $t_2$ a measurement is performed with an ideal-negative-result strategy: if the atom is in the state  $\ket{\!\!\downarrow}$ it experiences one field that displaces it far away and the corresponding result is discarded. In contrast, the atom in the state $\ket{\!\!\uparrow}$ does not experience this shift and it is consequently kept and allowed to further evolve according to the quantum walk until the final measurement at $t_3$; see \cref{fig:Robens}(b). From a macrorealist perspective, the atoms that interact with the field and are discarded are only those in the state  $\ket{\!\!\downarrow}$, so the measurement of the state $\ket{\!\!\uparrow}$ is performed in a noninvasive way.
The expression Robens {\it et al.} consider is
\begin{equation}\label{eq:Qwitn}
W:=\sum_{q_3 q_2} q_3 p(q_3 q_2) - \sum_{q_3} q_3 p(q_3) \leq 0 ,
\end{equation}
obtained by a specific choice of some trivial measurements in the three-term LGI. Interestingly, this expression closely resembles the inequality version of the NSIT condition. In addition, the authors also estimate the quantum witness $|W|$~\cite{LiSR2012}.

The authors evaluated the macroscopicity measure defined by
Nimmrichter and Hornberger in~\cite{Nimmrichter13}. Such a measure,
called $\mu$, sets a lower limit for the time (expressed in logarithmic scale) during which an electron is delocalized over distances larger than a certain length scale $\ell$, which represents a phenomenological parameter.
The value estimated in the experiment is $\mu = \log_{10}(T\hspace{2pt}M^2_a/m_\text{e}^2) \approx 6.8$, with $M_a$ and $m_\text{e}$ being the masses of the Cs atom and of an electron and $T$ being the total duration of the walk. The reference distance is thus set to
$\ell = 2\mu m$, which is the maximal distance along which the atom is displaced.

The measurement implemented here can indeed be called the ideal-negative result from a macrorealist perspective. However one needs to rely on the measurement model assumed by the experimenter, i.e., that the system interacts only with one of the two fields, depending on its internal state. Thus, the class of macrorealist theories ruled out by this experiment is of those describing the interaction between the field and the atom as a classical version of the quantum one.

\subsubsection{Single-photon excitation in macroscopically-separated crystals}
\begin{figure}[t]
	\includegraphics[width=\linewidth]{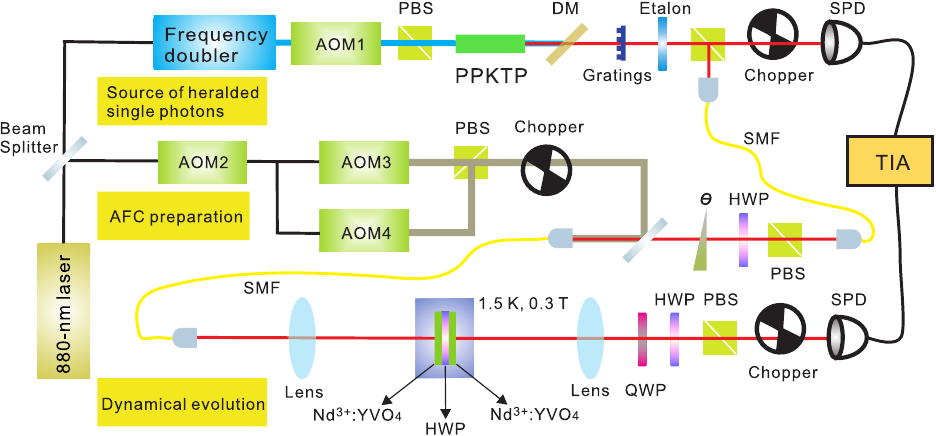}
	\caption{(Reproduced from \cite{ZhouPRL2015}). A delocalized atomic excitation is created between two pieces of Nd$^{3+}$:YVO$_4$ crystals which sandwich a $45^\circ$ half-wave plate (HWP). The evolution of the excitation is controlled
		with a polarization-dependent atomic-frequency comb	(AFC).
				 Photon pairs are generated in the PPKTP crystal and then spectrally filtered and separated by the polarization beam splitter (PBS). The H- and V-polarized photons are independently stored in the two pieces of crystals and acquire a different phase shift.  
				 After a programmable delay, the polarization of retrieved signal photons is analyzed with a quarter-wave plate (QWP), a HWP and a PBS. Single photon signals are detected with single-photon detectors (SPDs) and analyzed with time interval analyzers (TIA). 
	}\label{fig:Zhou}
\end{figure}

A different approach was pursued by Zhou {\it et al.} in Ref.~\cite{ZhouPRL2015}; see also Fig.~\ref{fig:Zhou}. Note that this experiment has been discussed also in the review \cite{EmaryRPP2014}. However, since then, the experiment has been updated, and especially further control runs have been made to test the Markovianity assumption on the two basis states. 
We recall that variants of the LGI based on Markovianity assumptions have been first derived in the context of quantum transport problems by Lambert {\it et al.}~\cite{LambertPRL2010}, and widely used for investigating macrorealism; see \cite{EmaryRPP2014} for more details.
In Ref.~\cite{ZhouPRL2015} a LGI-type inequality was tested, which relied on two assumptions different from the traditional macrorealism assumptions. The first assumption is {\it stationarity}, meaning that the conditional probabilities of finding the system in state $i$ at time $t+\tau$, conditioned on it being in state $j$ at time $t$, only depend on $\tau$, and the second assumption is {\it Markovianity}.

The systems consists of a single-photon atomic excitation displaced over two spatially separated crystals, whose dynamical evolution is controlled with a polarization-dependent atomic-frequency comb.
This system can be described as a qubit, using a collective Dicke state arising from the absorption of a photon. The two basis states are
$\ket{e}=\sum_{j=1}^N c_j e^{-ikz_j} e^{i2\pi \delta_j t} \ket{g_1,\dots,e_j,\dots,g_N}$ and $\ket{g}=\ket{g_1,\dots,g_j,\dots,g_N}$, where $N$ is the total number of atoms in the comb, and $\ket{e_j}$ and $\ket{g_j}$ are the excited and ground state of the atom in position $z_j$. The other parameters are the 
wave number $k$ of the input field, the detuning $\delta_j$ of the
atom with respect to the light frequency, and the amplitudes $c_j$ which depend on the frequency and on the position of atom $j$.

In the experiment, the system evolves with an oscillatory dynamics between these two basis states. Measurement performed in the qubit basis provide evidence that both stationarity and Markovianity hold. 
In particular, the authors measured at different times the four conditional probabilities ${p(i,t+\tau|j, t)}$ with $i,j=\{g , e\}$ referring to the two different states, ground and excited, and the evolution is an oscillatory dynamics between them. They also estimated the trace distance between the two basis states as it evolves in time, seeing that it decreases monotonically over time as it should for a Markovian dynamics.
Thus, in a sense they tested both assumptions with independent experiments.  However, they still rely strongly on  the assumption that only two basis states are involved in the dynamics.
Then, they tested the inequality $\mean{Q(0)Q(2t)}\pm 2\mean{Q(0)Q(t)} \geq -1$ by measuring over time the correlations of the dichotomic observable given by $\Q = \ketbra{g} - \ketbra{e}$, observing a sensible violation that follows perfectly the predictions of the quantum mechanical model.
Thus, given also the control tests performed, one could argue that the experiment disproves classical theories with two states.
Macroscopicity, as quantified by the disconnectivity, is very low ($D=1$) in the experiment, although the excitation was displaced between two macroscopic crystals separated by a length of the millimeter scale.
Finally, we remark that a subsequent experiment by the same group \cite{LiuPRA2019} managed to obtain a violation of the quantum witness constraint~\cite{LiSR2012}, in the form presented in \cite{KneeNC2016}, together with the additional experimental runs to estimate the invasivity of the measurement according to the scheme proposed in \cite{KneeNC2016} and discussed in more details in Sec.~\ref{sec:supercond} below. They obtained a superpositions of up to 76 atomic excitations shared by $10^{10}$ ions in two separated solids, corresponding to a disconnectivity of $D=5$.

\subsubsection{Neutrino oscillations}
A similar test of LG combined with the assumption of stationarity was performed on neutrino flavor oscillation~\cite{Formaggio2016}. One additional motivation for this test was that neutrinos can maintain coherence in very large distances due to their low interaction with the environment. In fact, neutrino oscillations were detected after a distance of approximately $735$ km in the data analyzed in \cite{Formaggio2016} and taken from the MINOS collaboration~\cite{MINOS}. 
In particular, violations of LG inequalities (with the assumption of stationarity) were detected
in oscillations between muon and electron-flavored neutrinos, i.e., between the two states $\ket{\nu_\mu}$ and $\ket{\nu_e}$ treated as a qubit system. After this initial experiment, several other tested violations of LGIs with two and three-flavored neutrino oscillations appeared~\cite{FuEPJ2017, GangopadhyayEPJ2017,  NaikooPRD2019}.
The assumption of stationarity as a substitute for noninvasivity was employed in order to sample identically prepared copies of the system rather than a single system evolving in time; technically, the authors analyzed the data at different frequencies, which correspond to different times. In fact, for the case of neutrino oscillations it is impossible to perform sequential measurement, so different assumptions, such as stationarity, are necessary. 

\subsubsection{Nuclear magnetic resonance}
Over recent years, a number of macrorealism tests have been performed using nuclear spin systems, such as nitrogen-vacancy centers or nuclear magnetic resonance (NMR) spectrometers. In particular, two recent experiments appeared 
\cite{KatiyarNJP2017,MajidyPRA2019} which used a Bruker DRX $700$MHz spectrometer, where the NMR samples are given by $^{13}C$ nuclei of a chloroform molecule, representing the system qubits, probed by $^1 H$ nuclei, playing the role of the ancillary qubits. A prototype dynamics for these experiments, as well as for other macrorealism tests, is given by a rotating spin, i.e., a system governed by the Hamiltonian $H_S = \Omega J_x$, where $J_x$ is the spin operator in a direction $x$ (not necessarily spin-$1/2$) and $\Omega$ is the rotation frequency (cf. \cref{sec:1A} and \cref{fig:simple_example}).
The measurement is performed along the $\Q=J_z$ direction through the ancillary spins and in these systems it is typically employed the ideal-negative-result scheme of neglecting the rounds in which a system-ancilla interaction 
is detected, which is meant to address the clumsiness loophole.
However, note that in liquid NMR samples another problem arises typically, which is the fact that the system is extremely noisy, with the preparation typically being $\varrho = \epsilon \ketbra{0} + (1-\epsilon) \id$ for $\epsilon\sim 10^{-5}$ or even smaller. This raises the issue of a fair-sampling assumption, which has to be made in such systems; see also Sec. 6 in Ref.~\cite{EmaryRPP2014} for a discussion about this point.
In \cite{KatiyarNJP2017} the experiment followed the idea of \cite{BudroniPRL2014} of testing macrorealism in multilevel systems and implemented the dynamics of a spin-$1$ system taken from the symmetric triplet subspace of two $^{13} C$ qubits.
In \cite{MajidyPRA2019} instead, they considered single qubits, but probed several versions of Leggett-Garg inequalities, especially regimes in which violation of macrorealism was detected with only subsets of them~\cite{HalliwellPRA2016}. 
Moreover, they employed the approach of \cite{HalliwellPRAContMeas}, in which a time-continuous measurement was performed of the {\it velocity} $v(t)=\dot \Q= \Omega J_y$ (so-called waiting detector).
With this measurement, they read off the time integrated quantity $\int_{t_i}^{t_j} v(t) {\mathrm d} t = \Q(t_i)- \Q(t_j)$ and could thus readily estimate the correlators from the second moments $\mean{(\Q(t_i)- \Q(t_j))^2}$.
As they point out, this measurement scheme allows us to substitute the noninvasivity assumption with that of the observable making only one change of sign~\cite{HalliwellPRAContMeas,MajidyPRA2019}. In the experiment, this was also complemented by an independent set of experiments with the negative-result measurements, in order to provide further evidence of negligible classical disturbance. 

\subsubsection{Superconducting qubits}\label{sec:supercond}
Several experiments have been also made with superconducting qubit systems. Knee {\it et al.}~\cite{KneeNC2016} performed  a macrorealism test that rules out classical models based on two-level systems. 
Concretely, they used a flux superconducting qubit and prepared it in an equal superposition of the two quantum states $\ket{\psi}=(\ket{0} + \ket{1})/\sqrt{2}$ and afterwards made a projective measurement in the computational basis. Then they performed some runs of the experiment with an additional blind measurement in between (i.e., a measurement without outputs) and evaluated the quantum witness $W$. To exclude classical theories larger than just macrorealism, they performed control experiments calculating the disturbance of the blind measurement in each of the two states $\{\ket{0},\ket{1}\}$ 
separately and subtracted from $W$ the value of the largest disturbance.
As a result, all interpretations based on classical mixtures of the two computational basis states are excluded, thus ruling out essentially a classical two-state model for their intermediate operation.

\begin{figure}[h!]
	\includegraphics[width=\linewidth]{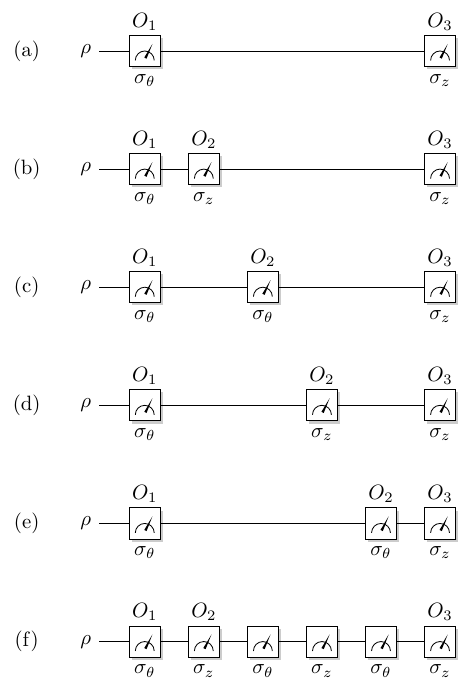}
	\caption{
		(Reproduced from \cite{HuffmanPRA2017}). The target qubit is prepared in an eigenstate of $\sigma_\theta$ for $\theta =-3\pi/4$, which corresponds to operation $O_1$. Protocol (a) is implemented by performing a measurement in the $\sigma_z$ eigenbasis after the preparation $\ketbra{1}_\theta$; for protocols (b) and (d) the state is transferred to a different ancillary qubit with a CNOT gate and then the ancilla is measured in the $\sigma_z$ eigenbasis; protocols (c) and (e) work similarly, just with the measurement preceded by a rotation of the state so that $\theta$ will be aligned to $z$. Finally, in protocol (f) all the measurements are implemented by sending sequentially the state of the target qubit to the four different ancillary qubits and making the measurements as in the corresponding protocols (b),(c),(d),(e).
			}\label{fig:Huffman}
\end{figure}

After this, a number of experiments with superconducting qubits have been implemented using the IBM 5Q Quantum Experience~\cite{HuffmanPRA2017,KU2020,SantiniVitale22}, which is a
system of five superconducting qubits that can be programmed via a publicly accessible website interface~\cite{IBM}.
In particular, Huffman and Mizel ~\cite{HuffmanPRA2017} implemented the protocol adroit-measurement protocol developed by Wilde and Mizel~\cite{WildeMizel2012}; see \cref{fig:Huffman}.

Concretely, the test was performed on one of the qubits, using the others as ancillas for implementing the 
protocol fulfilling the limitations and constraints of the IBM platform. The protocols (b), (c), (d), and (e), as 
described in \cref{fig:Huffman}, were used to estimate the adroitness of each intermediate measurement as 
$\epsilon_i=|\mean{O_3}_i-\mean{O_3}_a|$ where $i=\{b,c,d,e\}$ is a label indicating that the average is computed 
from the corresponding protocol containing one intermediate measurement.
The reasoning is then that a violation of the LG inequality 
$\mathcal L:=\mean{O_3}_a+\mean{O_2}_f+\mean{O_2 O_3}_f+1\geq 0$
with the additional verification that $|\mathcal L|\geq \epsilon_b+\epsilon_c+\epsilon_d + \epsilon_e$ signals a violation of macrorealism or some sort of collusion between the measurement disturbances.

An upgraded version of the same platform was used by Ku {\it et al.}~\cite{KU2020} to perform a macrorealism test preparing at time $t_1$ an optical cat states $\cos \tfrac \theta 2 \ket{0}^{\otimes n} + \sin \tfrac \theta 2 \ket{1}^{\otimes n}$ of $n=2,3,4,6$ qubits and then evolving it to the state $\ket{0}^{\otimes n}$ at time $t_2$. 
The idea is that these states have increasing disconnectivity for increasing $n$. 
The authors test the violation of an inequality of the form $|W| \leq \mathcal{I}(q=1)$, where the quantum witness is obtained as in \cref{eq:Qwitn} with the measured observable being $\Q = \ketbra{0}^{\otimes n}$. 
Here, the additional quantity on the right-hand gives a measure of the invasivity and is given by
$\mathcal{I}(q) = |1-p(q|q)|$, where $p(q_2|q_1)$ is the probability that the measurement changes the value of the observable from $q_1$ to $q_2$. This quantity has been independently measured in order to ensure a non-clumsy violation of macrorealism. 
In their protocol, the outcome $q=1$ is ideally obtained at $t_2$ with certainty if there are no measurements at $t_1$. However, when a measurement is performed at $t_1$ the ideal probability of observing outcome $1$ becomes $\sum_{q_1} p(q_2=1,q_1)=\cos^4 \tfrac \theta 2 + \sin^4 \tfrac \theta 2$. It is worth noticing here that ideally a higher difference is obtained with the state $\tfrac 1 {\sqrt 2^n}\left(\ket{0}+\ket{1}\right)^{\otimes n}$, which has minimal value of the disconnectivity (and in fact it is just a product state). With this, we emphasize once more the fact that a macroscopically entangled state is not strictly needed for the violation of macrorealism.

\subsubsection{Heralded single photons}
Finally, experiments were recently also conducted with heralded single photons. 
In the experiment discussed in Ref.~\cite{WangPRA2018}, the idea of Emary \cite{EmaryPRA2017} was implemented, namely, that of making ambiguous negative-result measurements on a three-level system. This allows for a replacement of the assumption of NIM with that of EIM, as discussed \cref{sec:loopholes}. 
Here, the three levels are encoded in different polarization/path states of one photon, and the measurements are performed to detect coincidences with a trigger photon, which is created in a pair with the target photon through parametric down-conversion. This also has a consequence that not all rounds are valid and thus a fair-sampling assumption has to be employed. In fact, notice that the ``measurement'' $\{\openone - \ketbra{q}\}_q$ is not a POVM, so it must necessarily be implemented via some postprocessing operation, in which the undesired results, in this case a no-detection event, are discarded. 

\begin{figure}[t]
	\includegraphics[width=\linewidth]{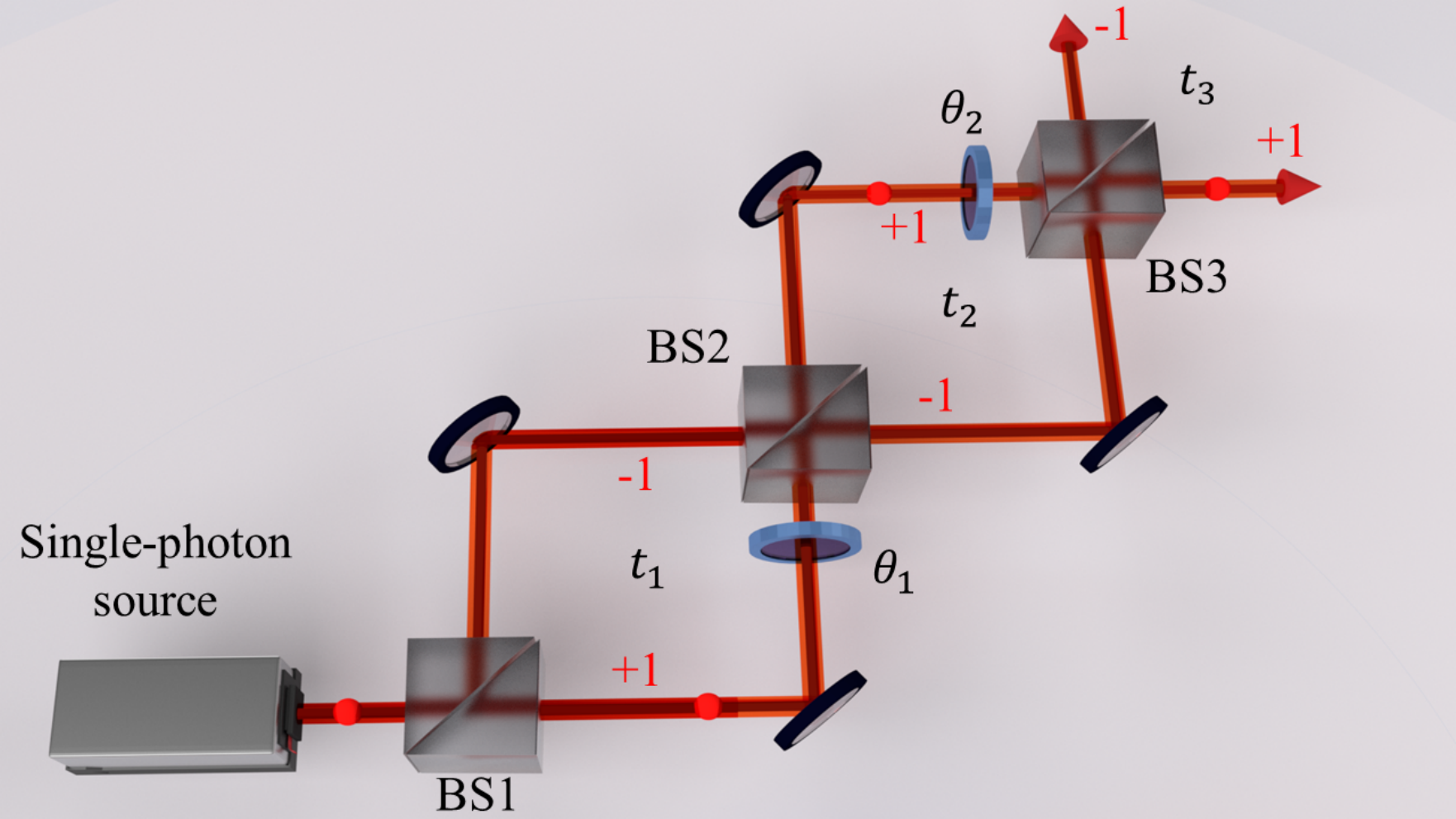}
	\caption{
		(Reproduced from \cite{JoarderPRX2022}). Schematic picture of the experimental setup of \cite{JoarderPRX2022}. The BS
		 are beam splitters, while the $\theta$ are phase modulators. As indicated in the figure, the two different paths in the interferometers correspond to the two different outcomes and the three time instants correspond to the paths after each BS.
	}\label{fig:Joarder}
\end{figure}

Joarder {\it et al.} \cite{JoarderPRX2022} also use heralded single photons, but perform a more traditional LG test;  see \cref{fig:Joarder}. This test has the merit of addressing very carefully the loopholes that arise from the detection of the events, in particular, problems such as the imperfect detection efficiency, the emission of multiple photons from the source or from different other sources, and the registration of coincidence events.
The clumsiness loophole here is addressed by performing negative-result measurements, and additionally ensuring that all the two-time NSIT conditions like \cref{eq:NSIT12} are satisfied. Thus, in this case the violation of LG inequalities comes from the genuine violation of a three-time NSIT condition. This is the case of two-adroit measurements that violate a LGI for sequences of three; see also the discussion in Sec.~\ref{sssec:WM_adroit}.
Furthermore, the argument for making classical explanations of the disturbance unlikely relies on the fact that measurements of the two different outcomes are performed in spatially well separated regions, which makes it an argument for the experiment to have a nontrivial degree of macroscopicity, in a similar sense as the experiment with neutrinos~\cite{Formaggio2016}. Note also that the particular case of macrorealism violation in interference experiments has been also discussed in more detail in other recent theoretical works~\cite{HalliwellEtAlPRA2021,Pan2020}.
Similarly as other experiments involving detections of photons (e.g., \cite{ZhouPRL2015}) or neutrinos \cite{Formaggio2016}, it was not possible to perform an actual sequence of measurements, but only a single (coincidence) measurement was performed at the end. The different times are then associated to longer paths taken by the photons (see \cref{fig:Joarder}).

To conclude this section, we note that experimental tests of LGI have significantly improved over the years: an increasing attention has been devoted to the analysis of possible loopholes, with the development of more refined theoretical arguments and experimental implementations. These approaches still rely on some description of the experimental apparatus and the measurement process and some additional assumptions that substitute the strong hypothesis of NIM. Nevertheless, one must admit that there are fundamental limitations in performing a LG test completely removing any assumption on the invasivity of the measurement. These efforts in the design and implementation of more elaborated experiments addressing the loopholes in LG tests contribute to disproving a larger set of MR models, thus providing increasingly stronger arguments against them.

\section{Nonclassical temporal correlations beyond noninvasive measurements}\label{sec:finitestates}

\subsection{Operational relaxations of noninvasive measurability}

Information processing tasks, classical or quantum, normally involve a series of steps where data is read from a memory, manipulated and written again. The sequentiality of such operations suggests that temporal correlations may play a role in explaining quantum advantages in information processing. However, Leggett-Garg NIM assumption imposes a strong restriction on which tasks can be analyzed: Any classical device with an internal memory that is updated sequentially would violate NIM. 
A similar drawback holds for the proposals that we have just analyzed to substitute the NIM assumption with an arguably weaker assumption on the measurement invasivity, which is then tested via control experiments.

In the light of this, a natural way to relax NIM assumption in a systematic way that is suitable for applications in information-theoretic tasks is to allow a bounded internal memory: The operations are allowed to be invasive, but they can modify an internal memory of at most $n$ bits; NIM is recovered for the special case $n=0$. This idea was first proposed by \.{Z}ukowski \cite{Zukowski2014} in the context of LGIs, and further explored in \cite{BudroniNJP2019}. 
This relaxation can be well formulated in information theoretic terms via a {\it finite-state machine} or {\it automaton} \cite{PazBook2003} a box that receives an input $s_1$, produces an output $q_1$, then  receives another input $s_2$ and produces another output $q_2$, and so on. It is described in terms of a classical probability as 
\begin{equation}
\begin{split}
p(\qon|\son)=\sum_{r_0,\ldots,r_n} p_0(r_0)p_{\rm T}(q_1,r_1,|r_0,s_1)\ldots\\
\ldots p_{\rm T}(q_n,r_n|r_{n-1} s_n),
\end{split}
\end{equation}
where $\{r_i\}$ are the internal states of the machines, $p_0$ their initial distribution, and $p_{\rm T}(q_{i},r_{i},|r_{i-1},s_{i})$ is the probability of emitting the output $q_i$ and transition to the internal state $r_i$, given that the internal state was $r_{i-1}$ and the measurement setting was $s_i$. This model corresponds to a classical model in which the system evolves through a sequence of internal states that are not observed and updated after every measurement. The number of internal states, called also the dimension or the memory of the system, is considered to be finite. 
Finally, notice that the transition rule $p_{\rm T}(q_{i},r_{i},|r_{i-1},s_{i})$ is fixed throughout the process, i.e., we have time-independent operations.

\begin{figure}
\includegraphics[width=\linewidth]{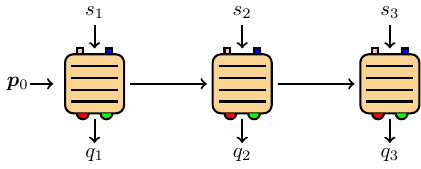}
	\caption{A device-independent temporal sequence. Boxes are either classical or quantum operations with input $s_i$ and output $q_i$ that act on a finite amount of states.
	The initial input can be taken as the vector representing, e.g, the first level $\vec p_0=(1,\dots,0)$.}
\end{figure}

When the dimension $d$ is allowed to go to infinity, the set of probabilities generated by such a model can cover the whole AoT polytope \cite{Hoffmann2018}, as it happens for quantum models. This is due to the fact that the extreme points of the AoT polytope are reached by deterministic strategies, as discussed in \cref{sssec:TC_gen}.
To implement such strategies, one needs sufficiently many states to store the whole past history at a given point in the sequence.

In contrast, when the number of states $d$ is kept constant, the extreme points of the AoT polytope, for given length and number of inputs and outputs, cannot in general  be reached.  Consequently, the sets of finite-dimensional quantum temporal correlations is strictly larger than the classical one. This fact can be explicitly witnessed by the violation of LG-type inequalities~\cite{BudroniNJP2019,BVWPRR2021,Vieira2021}. 
The memory resources needed to simulate the extreme points of the AoT polytope, i.e., the minimal number of internal states to reproduce the corresponding deterministic outcome-generation strategy via a classical model, was investigated by Spee {\it et al.}~\cite{SpeeNJP2020}, who showed that it scales at least exponentially in the length $n$.
In particular, a general criterion to compute such minimal dimension necessary to simulate the extreme points has been introduced in \cite{SpeeNJP2020}, based on the reuse of the internal states to generate an output sequence. It can be introduced via a basic example. Consider the sequence $010101$. It  has length 6, no inputs, and two possible outputs. It is clear that it can be generated with probability $1$ by a model with two internal states. Intuitively, the state at each time step fixes (deterministically) the future sequence, so to the same internal state it must correspond the same future. It is clear that in this case there are only two possible futures: (a) generate an alternating sequence starting from $0$ and (b) generate an alternating sequence starting from $1$. The general criterion then counts the number of {\it inequivalent futures} associated with an input-output sequence. In the case $n=2$ one obtains that the minimal dimension to reproduce all extreme points of the AoT polytope is given by $d=S+1$ where $S$ is the total number of settings \cite{SpeeNJP2020}.
This idea has been further developed into the notion of {\it deterministic complexity} \cite{Vieira2021} defined as the minimal number of internal states necessary to deterministically simulate a sequence $\mathbf{q}$ and denoted as ${\rm DC}(\mathbf{q})$. Differences between classical and quantum correlations, thus, arise for a sequence $\mathbf{q}$ only if $d< {\rm DC}(\mathbf{q})$. Moreover, DC can be efficiently computed in the case of sequences without input~\cite{Vieira2021}.

LG-type inequalities that bound classical and quantum models can then be derived by looking at AoT extreme points that cannot be reproduced deterministically. This was used in \cite{Hoffmann2018} to derive an inequality witnessing the quantum dimension. In \cite{BudroniNJP2019}, the case of different theories, i.e., classical, quantum, and {\it generalized probability theories} (GPTs) were investigated. For the two-input/two-output scenario one has
\begin{equation}\label{eq:firstdineq}
p(01|00) + p(10|10) + p(10|11) \leq \Omega_2^C < \Omega_2^Q < \Omega_2^{gbit} = 3 ,
\end{equation}
where the classical bound for a single bit $\Omega_2^C=9/4$ is strictly smaller than the bound for a single qubit $\Omega_2^Q\sim 2.35570$, and the gbit bound refers to a two-state GPT.

The assumption of a finite number of internal states has been used also in recent experiments~\cite{KneeNC2016,ZhouPRL2015} to substitute NIM with a weaker assumption.
In fact, those are also experiments with two inputs (i.e., ``measurement'' or ``no measurement'' at a certain time instant) and two outputs. 
However, in those experiments the two operations corresponding to the two different inputs are concretely given and independent control experiments were needed to exclude classical theories of a given dimension for those concrete operations; see also \cite{Schmid2022} for a comment on this assumption in the experiments~\cite{KneeNC2016,ZhouPRL2015}  and a proposed solution based on theory-agnostic tomography. Instead, inequalities like \eqref{eq:firstdineq} are {\it semi-device-independent} 
\cite{PawlowskiPRA2011}, namely, they assume nothing about the device except the dimension of the physical system. As such, a violation of the classical bound $\Omega_2^C$ would imply that there is no possible model for a single bit that can account for the observed probabilities.

In contrast to the standard spatial scenario, in the temporal one classical and quantum correlations can be distinguished even in the case of no inputs (or equivalently, just one input), due to the constraints of time-independent operations.
A family of inequalities, each valid for all classical automata of dimension $d$ is given by
\begin{equation}
	p(0\dots01) \leq \Omega^C_{d,n} ,
\end{equation}
where the sequence consists of $n-1$ $0$s and one $1$ at the end and the bound $\Omega^C_{d,n}$ depends on both $d$ and $n$.
Here, upper bounds on the value $\Omega^C_{d,n}$ can be computed for sequences up to length $20$ and a quantum model that violates this bound can be explicitly constructed~\cite{BVWPRR2021}. 
Intuitively this sequence, called the one-tick sequence in analogy with the problem of ticking clocks \cite{ErkerPRX2017}, is the one requiring the higher dimension to be reproduced deterministically. In fact, the system must ``count' that $n-1$ steps passed, then emit the output $1$. Such a deterministic model requires $n$ states. On the other hand, any sequence of length $n$ can be reproduced deterministically by a machine with $n$ states.

This sequence seems to play a special role among all sequences, as investigated in  \cite{Vieira2021}. The numerical optimizations performed on all sequences with no input and two outputs up to length 10, suggest that the maximum probability for a sequence $\mathbf{q}$ for a model of dimension $d$ is upper bounded by the maximum probability, over models of the same dimension, of the one-tick sequence of length 
${\rm DC}(\mathbf{q})$. In addition, the numerics also suggest the existence of a universal upper bound of $1/e$ for all sequences with no input and two outputs, whenever $d<{\rm DC}$.
In contrast, even a simple quantum model with a single Kraus operator can reach the algebraic bound $p(0\dots01)=1$ when $n={\rm DC}=d+1$ and we take the limit of both the length and the dimension going to infinity~\cite{Vieira2021}.

More generally, the idea of using some sort of memory complexity for quantifying the degree of nonclassicality of sequential protocols has appeared also in contexts independent from macrorealism. 
In particular, the concept of dimension witness, which has been first introduced in the context of Bell nonlocality~\cite{BrunnerPRL2008}, has been extended also to the temporal scenario. One of the first proposals involved a preparation and a (projective) measurement device~\cite{WehnerPRA08}, which has been extended in a number of ways, from a dimension witness based on the system's dynamics~\cite{WolfPerezGarciaPRL09}, to 
prepare-and-measure scenarios \cite{GallegoPRL2010, BowlesPRA2015}, contextuality  \cite{GuehnePRA2014} and sequential measurements \cite{Sohbi2021}, which stimulated several experimental tests~\cite{Ahrens2012,Hendrych2012,Ahrens2014,Spee_2020}. 

A notion of temporal correlations naturally arise in communication scenarios, such as quantum random access codes (QRACs)~\cite{Wiesner1983,Ambainis1999, Ambainis2002, BowlesPRA2015, AguilarPRL2018, Miklin2020}, communication complexity~\cite{Brassard2003, CC_review}, or communication cost for simulating Bell nonlocality \cite{Pironio_comm2003, Montina_comm2016,Ringbauer2017}. 
Within this framework, Brierly {\it et al.}~\cite{BrierleyPRL2015} introduced a notion of nonclassical temporal correlations. According to their definition, a temporal correlation $p(\mathbf{q}|\mathbf{s})$ generated by transmitting a $d$-dimensional quantum system through several parties that perform local measurements is nonclassical  if its classical simulation requires more than $\log(d)$ classical bits of communication. In other words, whenever a distribution can be generated by transmitting a $d$-dimensional quantum system, but not by transmitting a $d$-dimensional classical system, then such a correlation is nonclassical. To detect these nonclassical temporal correlations, the authors introduce a communication-complexity task called the sequential $n$-point modulo-$(m; d)$ problem. In this task, $n$ parties receive each some initial input, with a promise on the sum modulo $d$ of the inputs and they have to generate an output, with $m$ possible values, satisfying a condition on the total sum modulo $d$ of the outputs. They showed a gap in the amount of communication needed in the classical and quantum setup to solve the problem with probability $1$, thus, providing examples of nonclassical temporal correlations according to the definition given. Here the communication plays the role of the memory previously discussed: the message must be encoded in a classical or quantum memory that is then transmitted.
A similar idea appears also in a recent proposal to relax NIM assumption called retrievability of information (RoI)~\cite{Uola2022}. In simple terms, RoI assumes that the information ``disturbed'' by performing a first (invasive) measurement can be retrieved by adapting the second one depending on the outcome of the first.  In terms of a model satisfying MRPS (and not necessarily NIM) this assumption is precisely formulated as follows
\begin{equation}\label{eq:RoI}
\sum_{\lambda} p(\lambda) p(b|0,y, \lambda) = \sum_{\lambda,a} p(\lambda) p(a,b|x, y_a,\lambda), \ \forall y,x.
\end{equation}
If Eq.~\eqref{eq:RoI} holds, the corresponding MRPS model satisfies
\begin{equation}
p(b|0,y)=\sum_a p(a,b|x,y_a).
\end{equation}
Uola {\it et al.}~\cite{Uola2022} connect the violation of this condition to joint-measurability properties of the corresponding POVMs and the optimal retrieving protocols to the minimal value for the Busch-Lahti-Werner uncertainty relations~\cite{BLW_RMP_14}. We conclude this section by remarking that all the approaches that substitute NIM assumption with weaker one, e.g., limited memory, limited communication, or retrievable information, are all applicable to devise more and more stringent tests of macrorealism, even if they were not initially designed specifically for this goal.

\subsection{Applications of temporal correlations}\label{sec:applications}

We discussed how temporal correlations can be a mean of witnessing the minimal number of internal states, or  levels, of a physical system needed to describe a given experiment. This number can be interpreted as the memory, from a more information-theoretical perspective such as that of automata theory. The minimal amount of memory can be different for classical and quantum systems, giving rise to a notion of nonclassical temporal correlations and potential quantum advantages in sequential information processing tasks.
A typical question is the following: how much, in terms of classical memory, is the cost of a given task that needs sequential operations? This  was the motivation of some early works on the subject, including the more abstract introduction of the so-called ontological models~\cite{HarriganARXIV2007,GalvaoPRA2009}.

For quantum models, the question of the minimal Hilbert space dimension needed to explain some observed set of outcome probabilities $p(a|j,r)$, given preparation $r\in \{1,\dots , m\}$ and measurement input $j\in \{1,\dots,l\}$ was addressed first in \cite{WehnerPRA08}
making a parallel with the QRAC. Note that a similar question was also addressed in \cite{BrunnerPRL2008} and subsequent works in the context of spatial correlations. 

A different approach was followed in \cite{WolfPerezGarciaPRL09}, where the authors considered a (discretized) temporal evolution of expectation values of observables $\mean{A(t)}$ and, under the assumptions of Markovianity and time homogeneity, they proved a lower bound on the Hilbert space dimension $d$ given (i) the number of conserved quantities $D$ under a given evolution and (ii) the number $V$ of linearly independent sequences $v_t:=(\mean{A(t)},\mean{A(t+1),\dots})$ for $t\in \mathbb N$ (called delayed vectors). Such a bound is given by $D+V\leq d^2 +1$ and can be always attained by a quantum model. Remarkably, instead, a classical $d$-state model does not necessarily exists that saturates the bound.

Back to the static case, still in the prepare-and-measure scenario, classical and quantum dimension witness resembling the CHSH inequality were derived in \cite{GallegoPRL2010} with bounds $\Omega^C_d < \Omega^Q_d$ for the two cases; see also \cite{DellArnoPRA2012} for an analysis of their noise robustness and  ~\cite{Hendrych2012,Ahrens2012,Ahrens2014} for an implementation in photonic experiments. Later, other dimension witnesses beyond the prepare-and-measure scenario were derived~\cite{Spee2020,Sohbi2021,Mao_2022}. Recently, it has been also observed that dimension witnesses provide also lower bounds on the purity of the pre- and post-measurement states~\cite{SpeePRA2020}. 
Another early observation~\cite{KleinmannNJP2011} was that contextual models can be simulated via classical finite-state machines, having thus an associated memory cost. In particular, the case of the  Peres-Mermin square~\cite{Peres:1990PLA, Mermin:1990PRL}  was analyzed in detail \cite{KleinmannNJP2011, Fagundes2017}, showing a gap between the classical and quantum memory necessary to reproduce the temporal correlations. Moreover several noncontextuality tests involving sequential measurements were translated into proper dimension witnesses~\cite{GuehnePRA2014}.

A classical ontological model was proposed for quantum computation with magic states, a model for universal quantum computation~\cite{Zurel2020}. The operations allowed in the model are the state preparation, including preparation of magic states, Clifford gates, and Pauli measurements. The model can be easily mapped into a model based only on Pauli measurements. The classical model, then, is simply the model for a sequence of Pauli measurements on an $n$-qubit state. The authors essentially find a classical finite-state machine representation for such measurement sequence, as it was to be expected from the result of \cref{sssec:TC_gen}. Such a model, however, requires a classical state space with dimension scaling exponentially with respect to the number of qubits. It is possible that such a model can be improved, as no argument of optimality was provided.

Temporal correlations were shown to play a central role in the performance of time-keeping devices~\cite{ErkerPRX2017,BVWPRR2021,Woods2022}.
Specifically, one can model a clock as a device that ticks at regular time intervals according to a distribution $p(t)$, i.e., the probability of ticking at a background time value $t$.
In this model, a notion of {\it accuracy} can be given as $R=\mu^2/\sigma^2$ where $\mu:=\sum_{t=1}^\infty  t p(t)$ is the mean and $\sigma^2 := \sum_{t=1}^\infty (t-\mu)^2 p(t)$ is the variance. $R$ can be interpreted as the average number of ticks after which the clock uncertainty is greater than the interval between two ticks. It has been observed that $R$ is bounded by $O(d)$ and by $O(d^2)$ for, respectively classical and quantum $d$-state models~\cite{BVWPRR2021,Woods2022}. Here we use a notation for a discrete-time clock, but the continuous limit has been also considered~\cite{BVWPRR2021,Woods2022}.
Furthermore, optimal or close-to-optimal classical and quantum models have been investigated for this task, which also triggered subsequent investigations into dimension witnesses based on sequences without inputs~\cite{Vieira2021}. Note, in fact, that such a ticking-clock model can be seen as a machine that has no input (i.e., it is repeated equally at each time step) and two outputs (tick and no-tick).
The investigation of quantum clocks naturally involves discussions about thermodynamical irreversibility and entropy production~\cite{ErkerPRX2017}. In addition to that, temporal correlations for system of finite-size were  also investigated~\cite{HenautPRA2018,Catani2022} from the perspective of irreversibility and Landauer's principle~\cite{Landauer61}, showing that the requirement of reversible transformations limits the amount of temporal correlations obtainable from a quantum system; see also \cite{ReebNJP2014,Tarantoetal} for a discussion of finite-size effects on Landauer's principle.

\section{Outlook and challenges for the future}\label{sec:outlook}

Let us conclude this perspective paper by discussing the possible future directions of the research on Leggett-Garg macrorealism and temporal quantum correlations.
We have seen that the notion of macrorealism arises as a hidden-variable model in the temporal scenario, in analogy with the early works of Bell and Kochen-Specker on local realism and contextuality. As such, many of the techniques developed for these fields, like the correlation polytope approach to the computation of Bell and noncontextuality inequalities, can be applied also to macrorealism. However, MR models have a very peculiar property, namely the possibility of measuring directly the global probability distribution from which all possible observations should arise as marginals. As such, the problem can be straightforwardly solved by checking the NSIT conditions, unless additional limitations are imposed on the sequences of measurements that can be performed; see \cref{sec:1}. The question remains open of what  the optimal tests are, not only with respect to MR models, but also with respect to quantum violations of LGIs and possibly additional figures of merit, such as quantitative estimates of measurement disturbance and more convoluted measurement procedures that appear when trying to close the loopholes. In particular, in future experiments it would make sense to consider measurements which are more general than projective, and potentially also with different state-transformation rules, and to make model-independent tests, which have the weakest possible assumptions.

Indeed, Leggett-Garg tests are subject to several loopholes, as it happens to Bell and contextuality tests. The hardest condition to verify experimentally is the fact that the measurement is noninvasive. Performing a LG test without the NIM assumption may indeed be fundamentally impossible. However, what can be done and what has been done in experiments is to substitute NIM with some weaker assumption that can be supported by additional experimental evidence. This approach may not provide a complete loophole-free experiment, but it can certainly disprove an increasingly large class of MR models. Recent years have seen a considerable progress in this direction, both from the perspective of theoretical arguments and experimental implementations; see \cref{Sec:exp_tests}. 

A systematic approach to the relaxation of the NIM assumption is to use the notion of finite-state machines~\cite{Hoffmann2018,BudroniNJP2019,SpeeNJP2020}, i.e., automata with a finite number of internal states. 
This approach provides a hierarchical and mathematically well formulated set of conditions able to distinguish between classical and quantum correlations. With this notion, MR models are recovered as a machine with only one internal state.  A natural relaxation of this condition consists of models with an increasing number of internal states, which, in the limit of an infinite number of them, are able to simulate all possible temporal correlations; see \cref{sec:finitestates}.
 Some of the LG experiments already went in this direction, by disproving with extra control experiments a classical model with the same number of states as the quantum model. In other words, the classical model is analyzed under the assumption of a finite-number of internal states. Typically a classical model with two states is compared with a qubit experiment~\cite{KneeNC2016,LiuPRA2019}. However, these experiments using control tests of invasivity only disprove a particular classical model, i.e., they are not model-independent. The assumption of finite system size, described in Sec.~\ref{sec:finitestates}, provides a systematic approach to address the loopholes in LGI tests by replacing the NIM assumption with a weaker one, namely, that of finite system size. Moreover, as discussed in Sec.~\ref{sec:finitestates} the corresponding temporal inequalities are independent of the particular classical model chosen, but depend only on the system size. Future experiments may consider this direction for tightening the loopholes in tests of macrorealism and disproving an increasing number of MR models.

Inequalities that are valid for all of either classical or quantum models with a certain number of states have been derived theoretically and also tested in some experiments with the purpose of witnessing the dimension of the state space, in a way that is independent of any macrorealist consideration.
This kind of approach has also the advantage of finding more concrete applications in other quantum information tasks; see \cref{sec:applications}).

The problem of the characterization of temporal correlations and sequential quantum information processing tasks can be addressed in the more general framework of spatiotemporal correlations. For instance, spatiotemporal correlations can be described via the {\it quantum comb} formalism \cite{ChiribellaDArianoPerinottiPRA2009}, when events have a well-defined temporal order, or the process matrix formalism ~\cite{Oreshkov_2012,ChiribellaDArianoPerinottiValironPRA2013} when they do not. These approaches try to capture a general notion of quantum process that involves a multi-partite and multi-time scenario, and in general can weaken the assumption of causality, intended as a definite time-ordering of the events. See also \cite{CostaPRA2018} for an approach that tries to unify the notion of spatial and temporal quantum correlations within this framework. Similarly, MR models can be generalized to more complex sets of causal relations, described by Bayesian networks~\cite{Pearlbook}. Some recent research in this direction attempts to relax some of the LG assumptions and substitute them with some limitations on the causal influences between classical variables~\cite{Ringbauer2017}.

Furthermore, we mention that temporal correlations are intimately connected to the problem of non-Markovian dynamics. Several works explored these connections \cite{ChenPRL2016,MilzPRX2020} and assumptions of Markovianity play a role in experimental tests of macrorealism \cite{ZhouPRL2015}. Similar assumptions are also used in the context of dimension witnesses \cite{WolfPerezGarciaPRL09} .
These results elucidate to some extent the connections between Markovianity, supplemented by stationarity (i.e., loosely speaking time-translation invariance) and the original LG formulation of macrorealism. Moreover, they can be similarly related to finite-state automata with just a single state. At the same time, the investigations around the notion of Markovianity are far wider~\cite{RivasRev2014,BreuerRev16,Milz_2021}
and the precise relation with macrorealism has not been explored in full detail, especially in the light of the latest developments in both fields. 
As a potential outlook, the concept of Markovianity can turn out to be crucial also for extending the notion of temporal hidden variable theories to the case of systems with an infinite number of internal state. Similarly, one can look at a definition of automata with an unlimited number of states. In other words, the notion of finite memory, in terms of a state space ($n$ states), could be substituted with the notion of a finite memory in terms of a time interval ($n$ time-steps). In this scenario, it is still an open challenge to make a meaningful and perhaps operational distinction between classical and quantum models for a sequential scenario. 

Another important challenge is to explicitly characterize temporal correlations as resources for specific quantum information processing tasks, such as phase estimations, quantum computation or the classical simulation of a sequence of quantum operations. A natural question, for instance, is whether the classical model of Zurel {\it et al.}~\cite{Zurel2020} can be optimized to further reduce the number of classical states necessary for the simulation of quantum computation and what is the minimal number. An answer to this question may have important consequences for our understanding of quantum advantages in computation.

Finally, these quantum information applications provide a further motivation for conducting increasingly elaborated LG tests. The finer control of quantum systems over many degrees of freedom, possibly even at some macroscopic level, and over sequences of operations, developed to address the loopholes in LG tests, may turn out to be crucial for the realization of sequential information processing tasks inspired by the study of temporal correlations.  At the same time, efforts in that direction are also required to tackle further foundational questions, as well as for new technological developments, such as better sensors, communication networks or quantum computers.

\acknowledgements{ 
The authors thank Shin-Liang Chen, Huan-Yu Ku, David Schmid, and Roope Uola for useful comments on the paper. This work is supported by the Austrian Science Fund (FWF)
through Projects  P 35810 and  P 36633
(Stand-Alone),   ZK 3 (Zukunftskolleg), and  F 7113
(BeyondC). 
}

\bibliography{biblio}{}

\end{document}